\documentclass[a4paper,fleqn,usenatbib,useAMS]{mnras}

\usepackage{graphicx}
\usepackage{amsmath}
\usepackage{amssymb}
\usepackage{multicol}        
\usepackage{bm}
\usepackage{pdflscape}
\usepackage[encapsulated]{CJK}
\usepackage{ucs}
\usepackage[utf8x]{inputenc}

\usepackage[dvipsnames]{xcolor}
\usepackage{tikz,hyperref}

\definecolor{lime}{HTML}{A6CE39}
\DeclareRobustCommand{\orcidicon}{
  \begin{tikzpicture}
    \draw[lime, fill=lime] (0,0) 
    circle [radius=0.14] 
    node[white] {{\fontfamily{qag}\selectfont \tiny ID}};
    \draw[white, fill=white] (-0.0625,0.095) 
    circle [radius=0.007];
    \end{tikzpicture}
  \hspace{-2mm}
}

\foreach \x in {A, ..., Z}{\expandafter\xdef\csname orcid\x\endcsname{\noexpand\href{https://orcid.org/\csname orcidauthor\x\endcsname}
    {\noexpand\orcidicon}}
}

\usepackage[T1]{fontenc}
\usepackage{ae,aecompl}

\usepackage{txfonts}

\title[Dust properties of RCW\,58]{Dust in RCW\,58: clues to common envelope channel formation?}
\author[Jim\'{e}nez-Hern\'{a}ndez et al.]{P.\,Jim\'{e}nez-Hern\'{a}ndez\thanks{E-mail:\,p.jimenez@irya.unam.mx}$^{1\orcidA}$, S.\,J.\,Arthur$^{1\orcidB}$, J.\,A.\,Toal\'{a}$^{1\orcidC}$ and A.P.\,Marston$^{2}$\\
$^{1}$Instituto de Radioastronom\'{i}a y Astrof\'{i}sica (IRyA), UNAM Campus Morelia, Apartado postal 3-72, 58090 Morelia, Michoacan, Mexico\\
$^{2}$European Space Astronomy Centre, ESA, PO Box 78, 28691 Villanueva de la Ca\~{n}ada, Madrid, Spain
}

\pubyear{2021}

\begin{document}
\label{firstpage}
\pagerange{\pageref{firstpage}--\pageref{lastpage}}
\maketitle

\begin{abstract}
We present a characterization of the dust in the Wolf-Rayet (WR) nebula RCW\,58 around the WN8h star WR\,40 using archival infrared (IR) observations from {\it WISE} and {\it Herschel} and radio observations from ATCA. We selected two clumps, free from contamination from material along the line of sight and located towards southern regions in RCW\,58, as representative of the general properties of this WR nebula. Their optical, IR and radio properties are then modelled using the photoionization code {\sc cloudy}, which calculates a self-consistent spatial distribution of dust and gas properties. Two populations of dust grains are required to model the IR SED: a population of small grains with sizes 0.002--0.01~$\mu$m, which is found throughout the clumps, and a population of large grains, with sizes up to 0.9~$\mu$m, located further from the star. Moreover, the clumps have very high dust-to-gas ratios, which present a challenge for their origin. Our model supports the hypothesis that RCW\,58 is distributed in a ring-like structure rather than a shell, and we estimate a mass of $\sim$2.5~M$_\odot$. This suggests that the mass of the progenitor of WR\,40 was about $\approx40^{+2}_{-3}$~M$_\odot$. The ring morphology, low nebular mass, large dust grain size and high dust-to-gas ratio lead us to propose that RCW\,58 has formed through a common envelope channel, similar to what has been proposed for M\,1-67. 
\end{abstract}





\section{INTRODUCTION}
\label{sec:intro}

Wolf-Rayet (WR) stars are identified by their broad emission lines and classified 
 according to the presence of strong helium and nitrogen (WN subtypes), carbon (WC) or oxygen (WO) lines in their spectra \citep[see][and references therein] {Crowther2007}.

Between O supergiants and WN stars there is a continuity of physical properties, and WN stars with hydrogen emission lines in their spectra (denoted WNh) have been proposed to be distinct from classical H-free Wolf-Rayet stars. In a single massive star evolutionary scenario, WC and WO subtypes represent the most advanced stages of the WR phenomenon just prior to explosion as Type Ibc supernovae. Classical WN stars are core helium-burning stars, while WNh stars are thought to be in the final stages of core hydrogen burning \citep{Hamann2006,Smith2008}.

Late-type WN stars, in particular those of the WN8h sub-type, 
often have peculiar characteristics:  
they are among the most massive and luminous WR stars, exhibit optical variability, and count several runaway stars among their number \citep[e.g.,][]{Moffat1986,Moffat1989,Antokhin1995,Marchenko1998}. Such characteristics have led different authors to suggest that they harbour hidden companions \citep[see][and references therein]{Toala2018} and   even that they are related to the exotic Thorne-\.{Z}ytkow objects \citep{Foellmi2002}.

In \citet[][hereinafter Paper~I]{Palmira2020}, we began a series of studies of the gas and dust in WR nebulae surrounding WN8h stars. Nebulae retain the imprint of the previous evolution history of their central star and its various mass-loss stages. Our first target was the iconic nebula M\,1-67 around WR\,124. This ejecta nebula shows no evidence of interaction with the surrounding ISM due to its location high above the Galactic plane. We used the photoionization code {\sc cloudy} \citep{Ferland2017} to model multi-mission infrared (IR) observations in addition to previously published optical spectroscopy. For nebulae around hot stars, it is important to model the photoionized gas and dust self-consistently, because both components absorb EUV photons, whereas only the dust absorbs FUV photons. Our {\sc cloudy} model of M\,1-67 successfully reproduced the IR thermal dust emission and the main optical emission lines and points to the presence of silicate grains with sizes as large as 0.9~$\mu$m and a dust-to-gas mass ratio of 4~per~cent.  \citet{Kochanek2011} showed that formation of large dust grains in the winds of evolved massive stars most likely occurs through eruptive events because the mass-loss rate needs to be high enough to shield
the dust-formation region from soft UV photons. To obtain dust grains as large as 0.9~$\mu$m requires mass-loss rates $\dot{M} > 10^{-3}$~M$_\odot$~yr$^{-1}$. 
Additionally, we found a bipolar distribution of the thermal dust emission, which supports a binary system origin.  
We suggested that  M\,1-67 was formed by the ejection of the H-rich envelope of the progenitor of WR\,124 in a common envelope event produced by an unseen companion. This would make M\,1-67 the first observational evidence of the formation of a WR nebula through a common envelope channel.

\begin{figure*}
\centering
\includegraphics[width=0.32\textwidth]{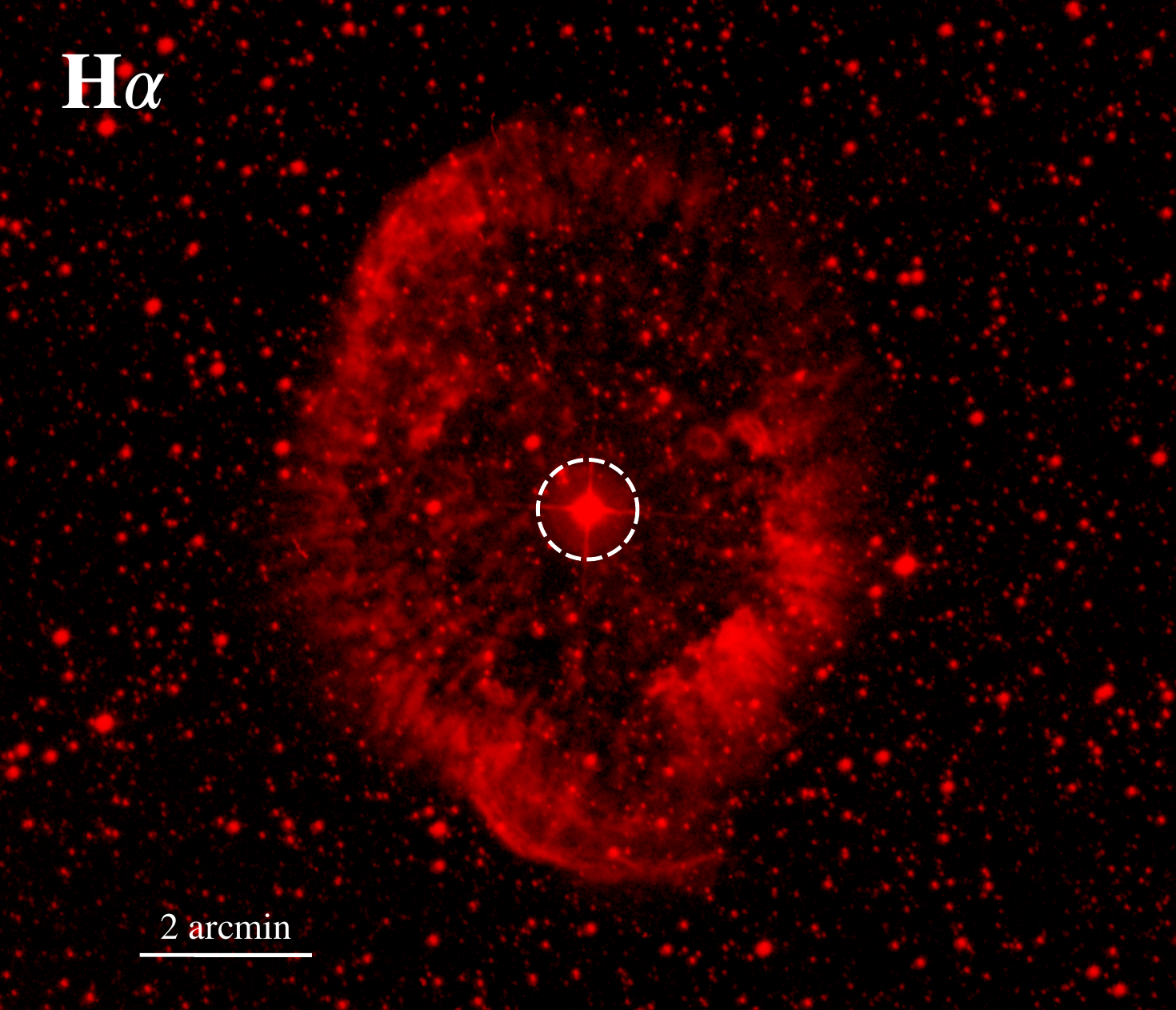}
\includegraphics[width=0.32\textwidth]{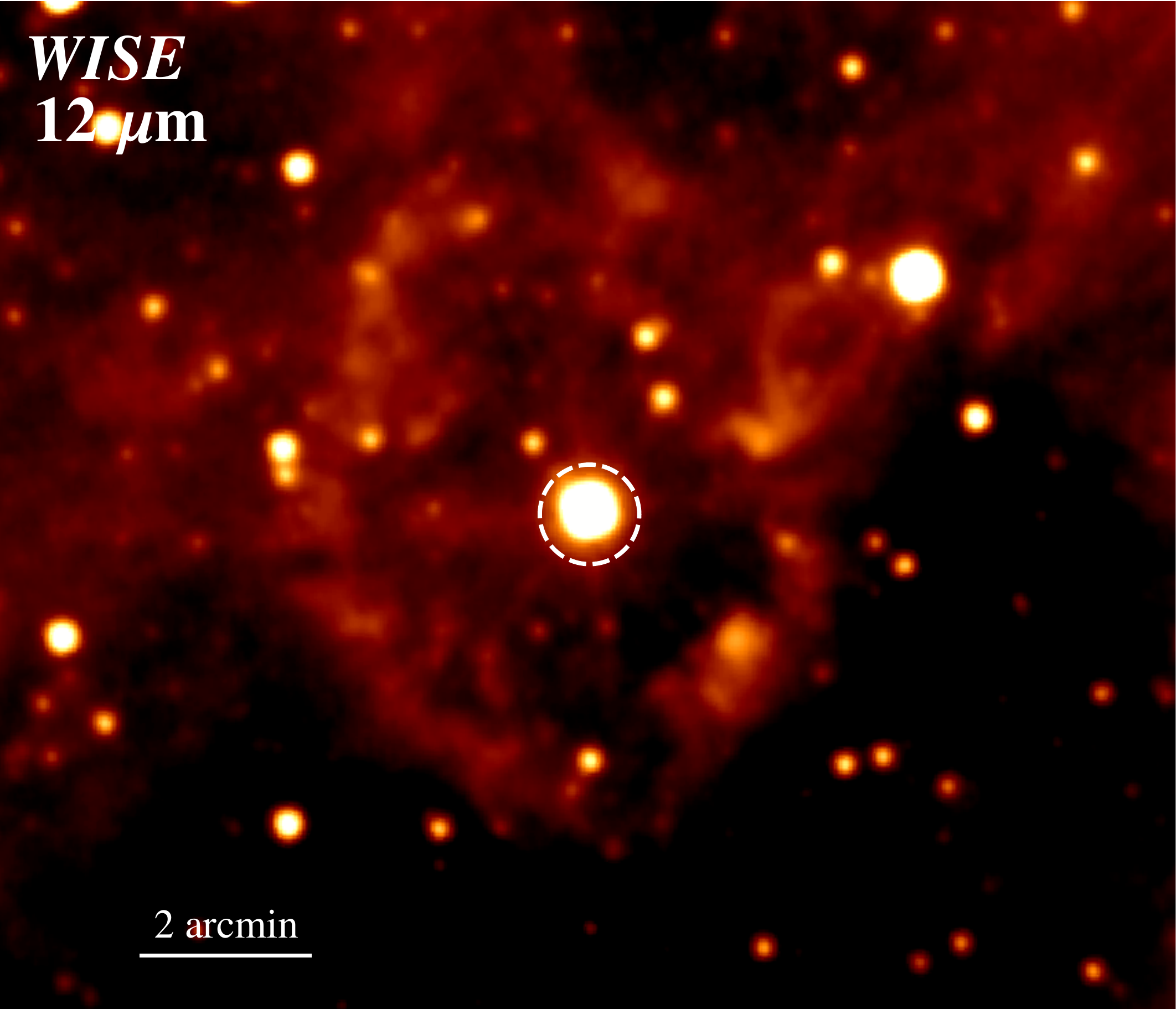}
\includegraphics[width=0.32\textwidth]{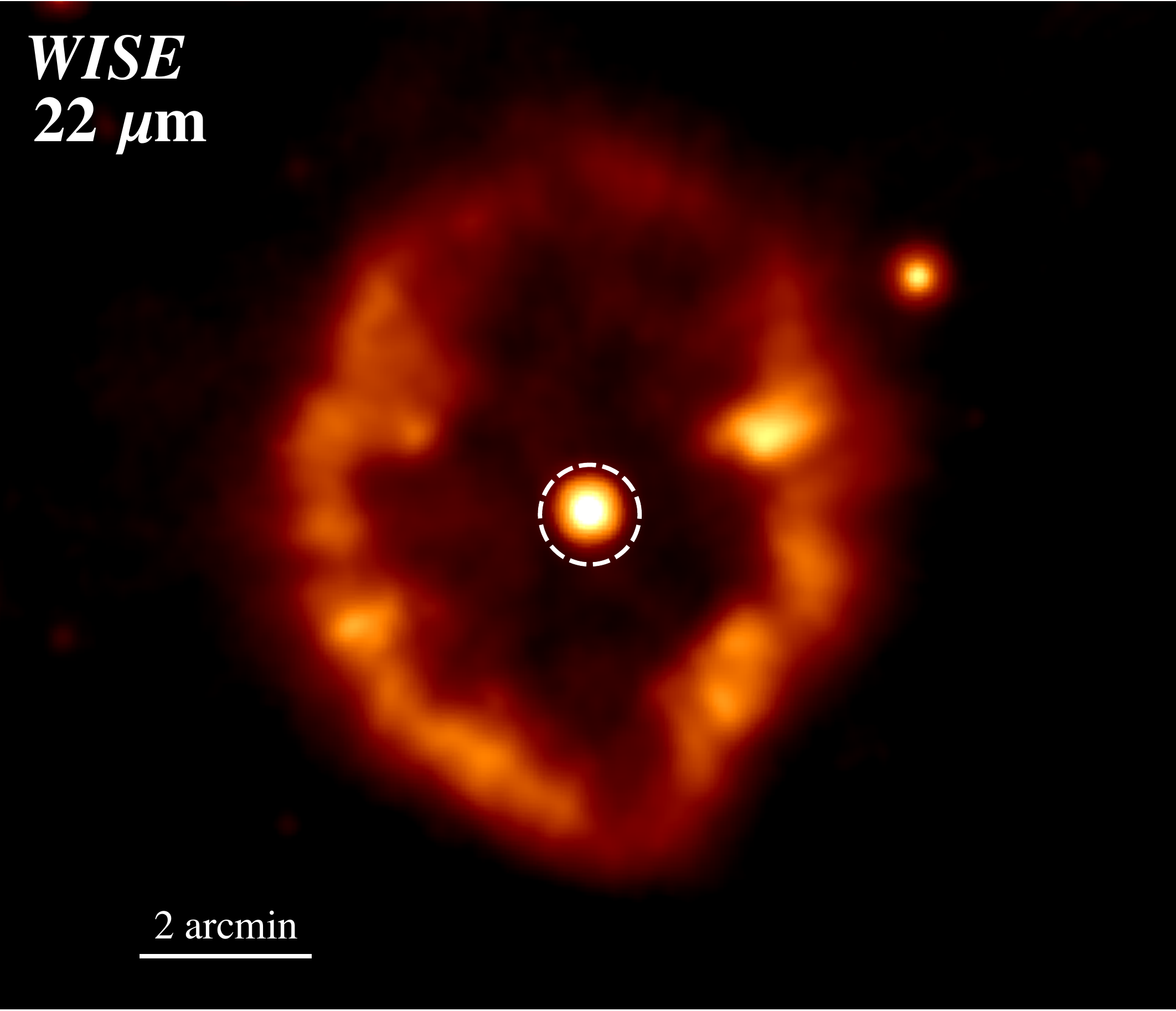}\\
\includegraphics[width=0.32\textwidth]{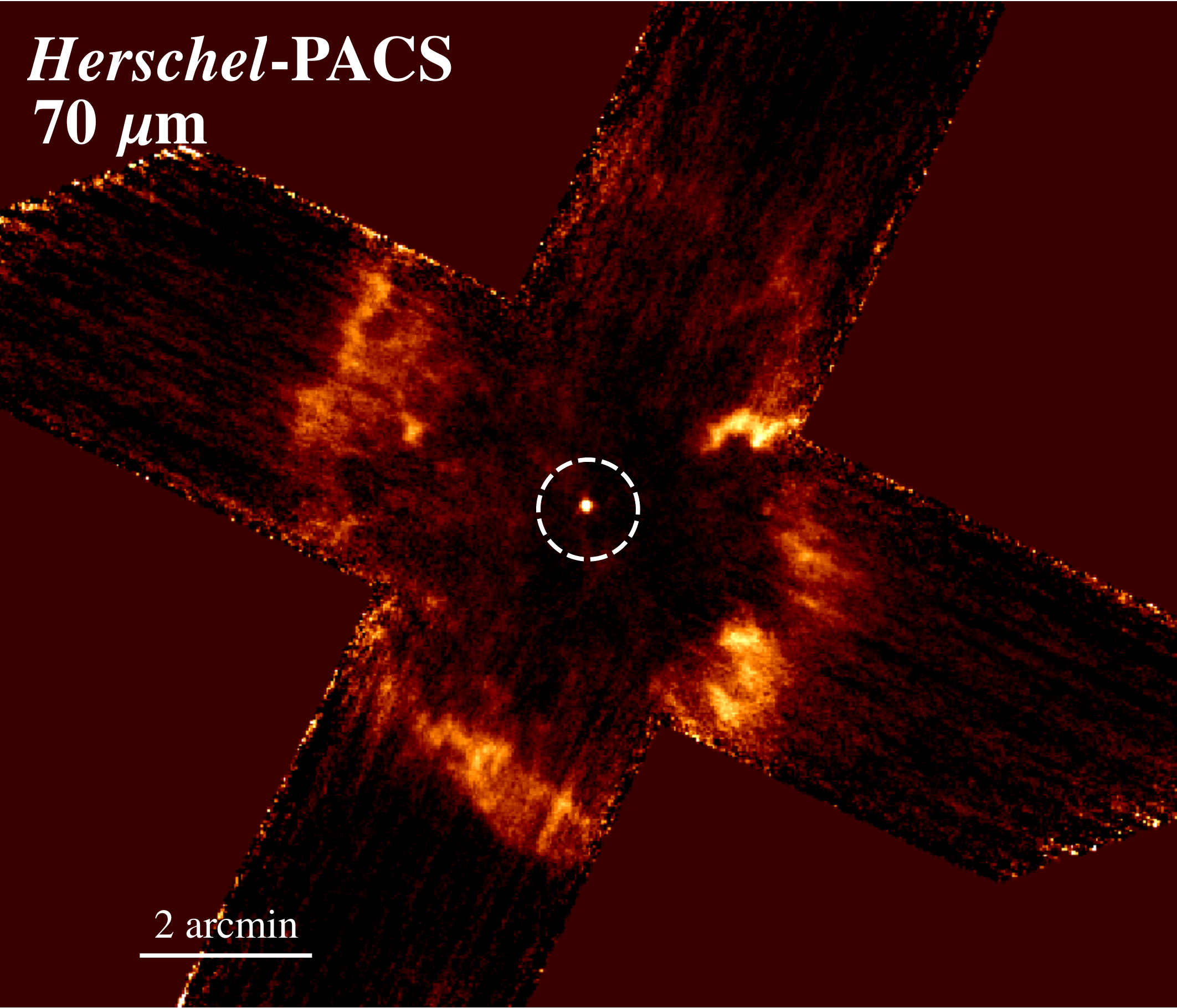}
\includegraphics[width=0.32\textwidth]{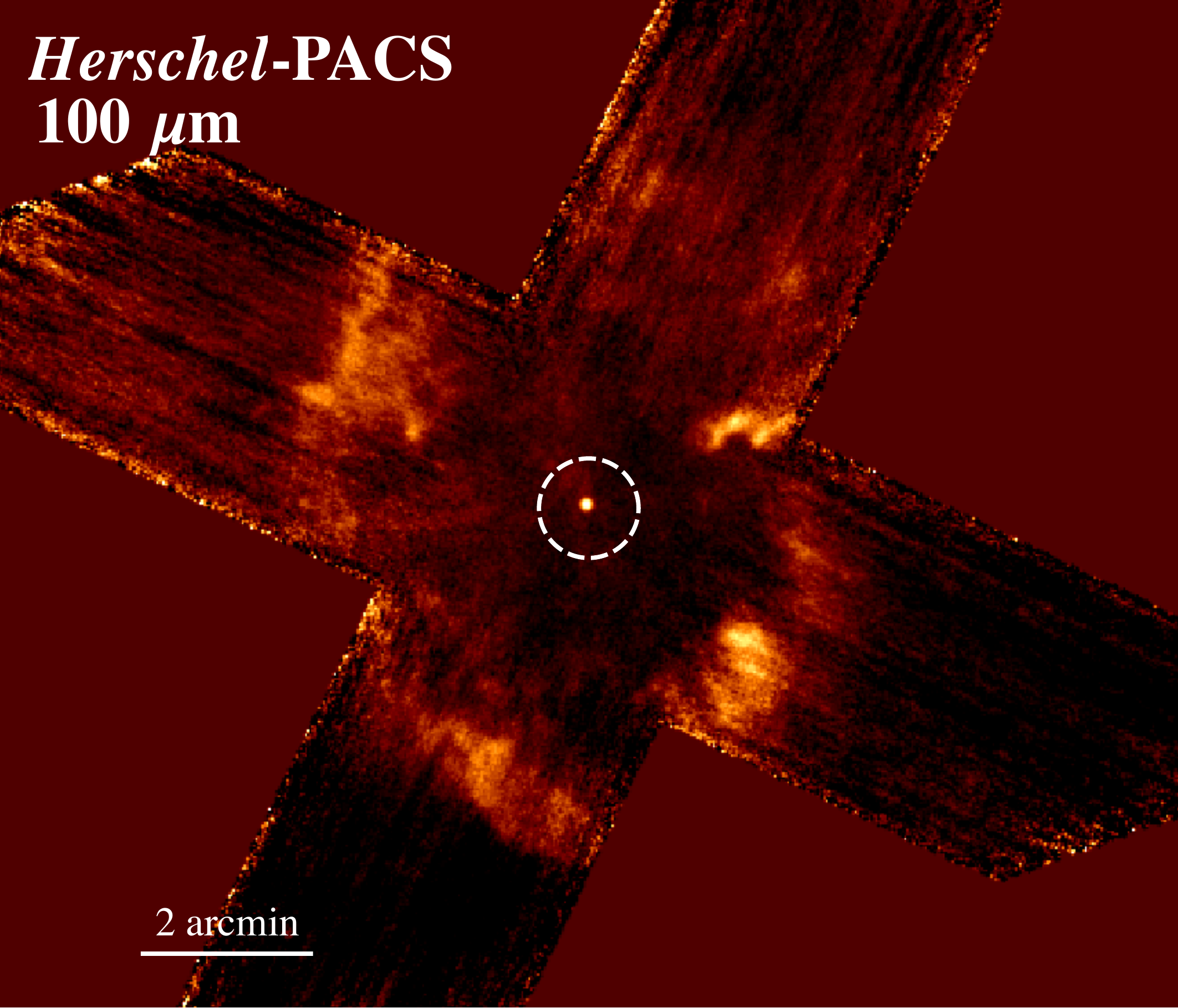}
\includegraphics[width=0.32\textwidth]{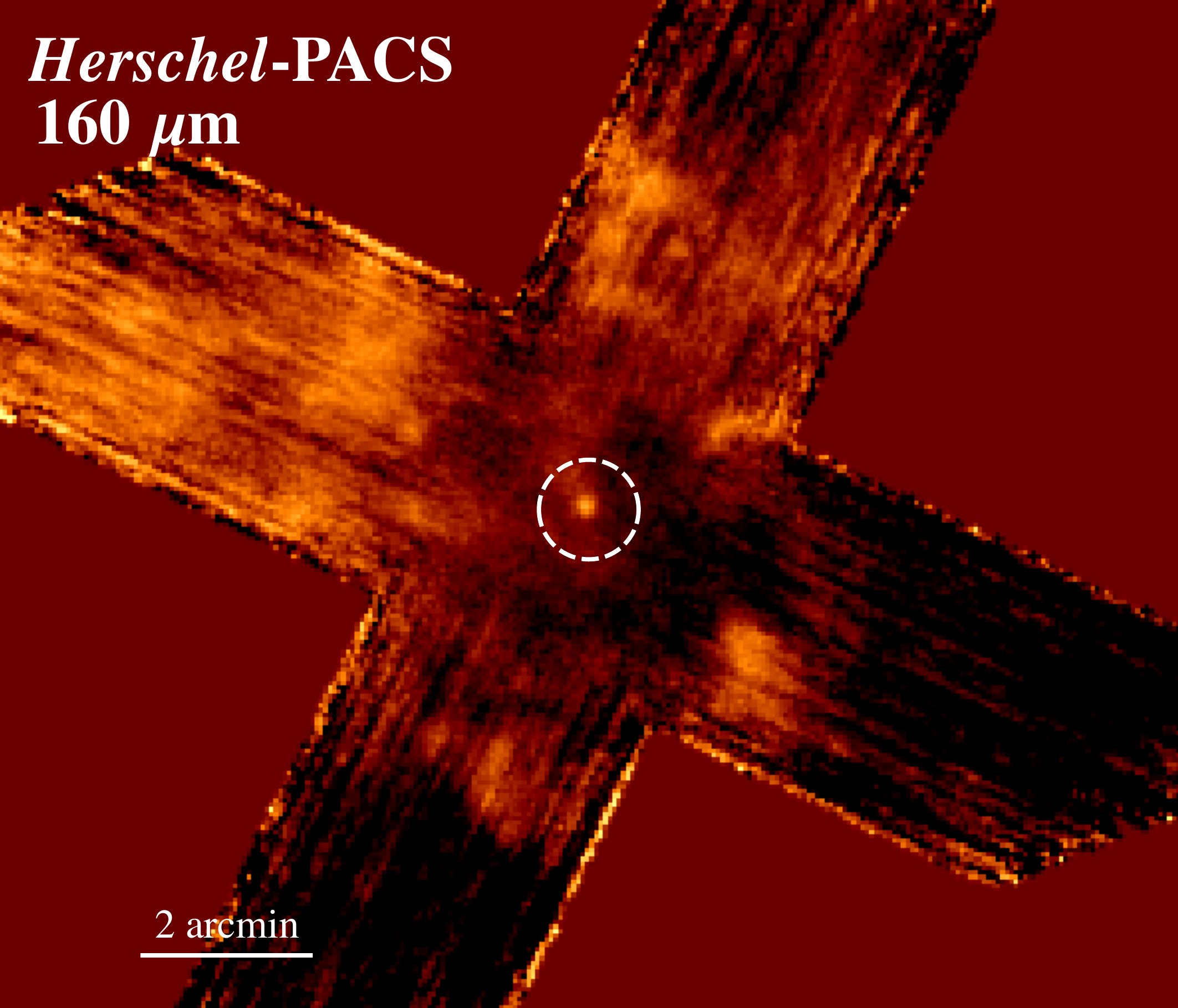}\\
\includegraphics[width=0.32\textwidth]{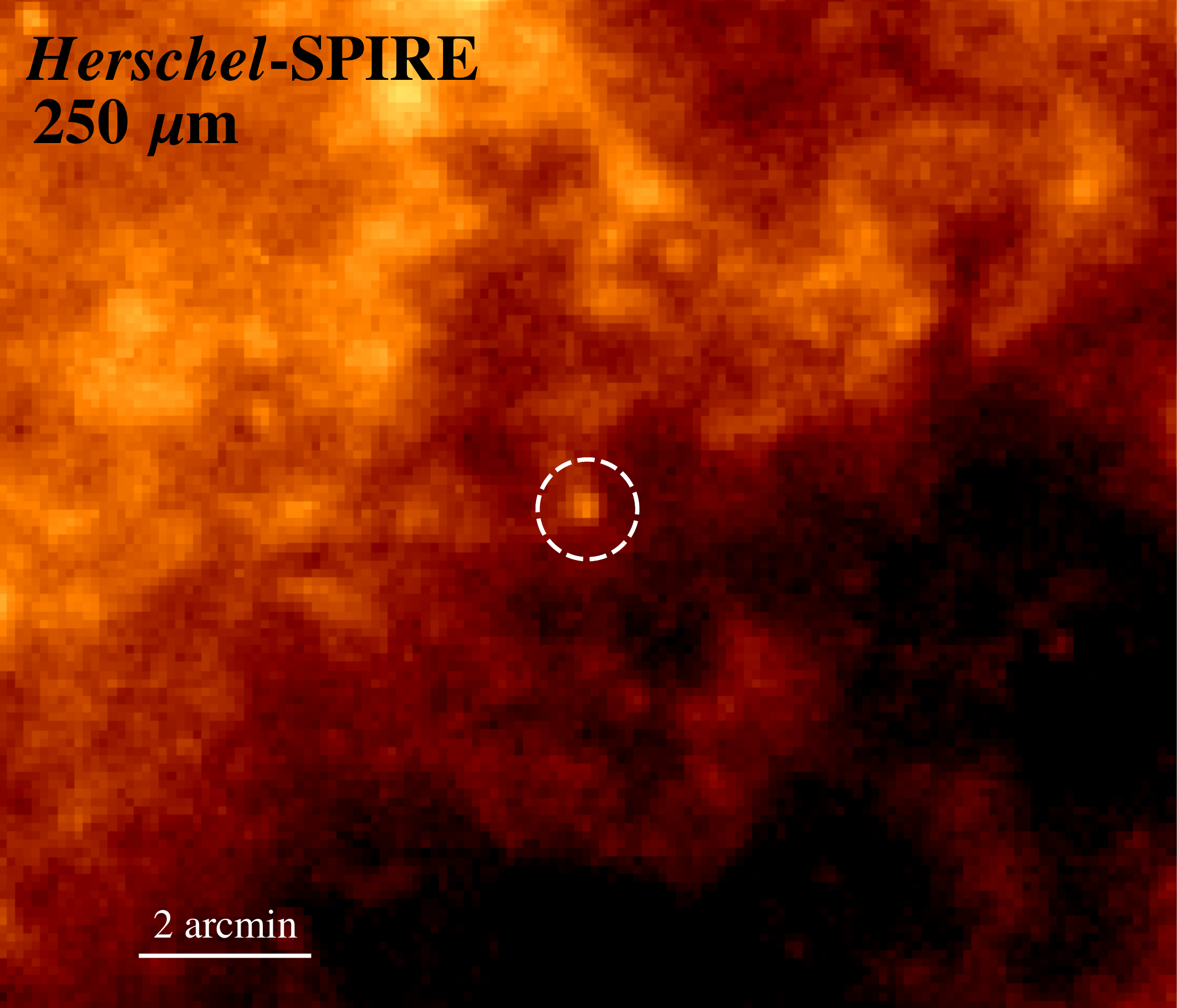}
\includegraphics[width=0.32\textwidth]{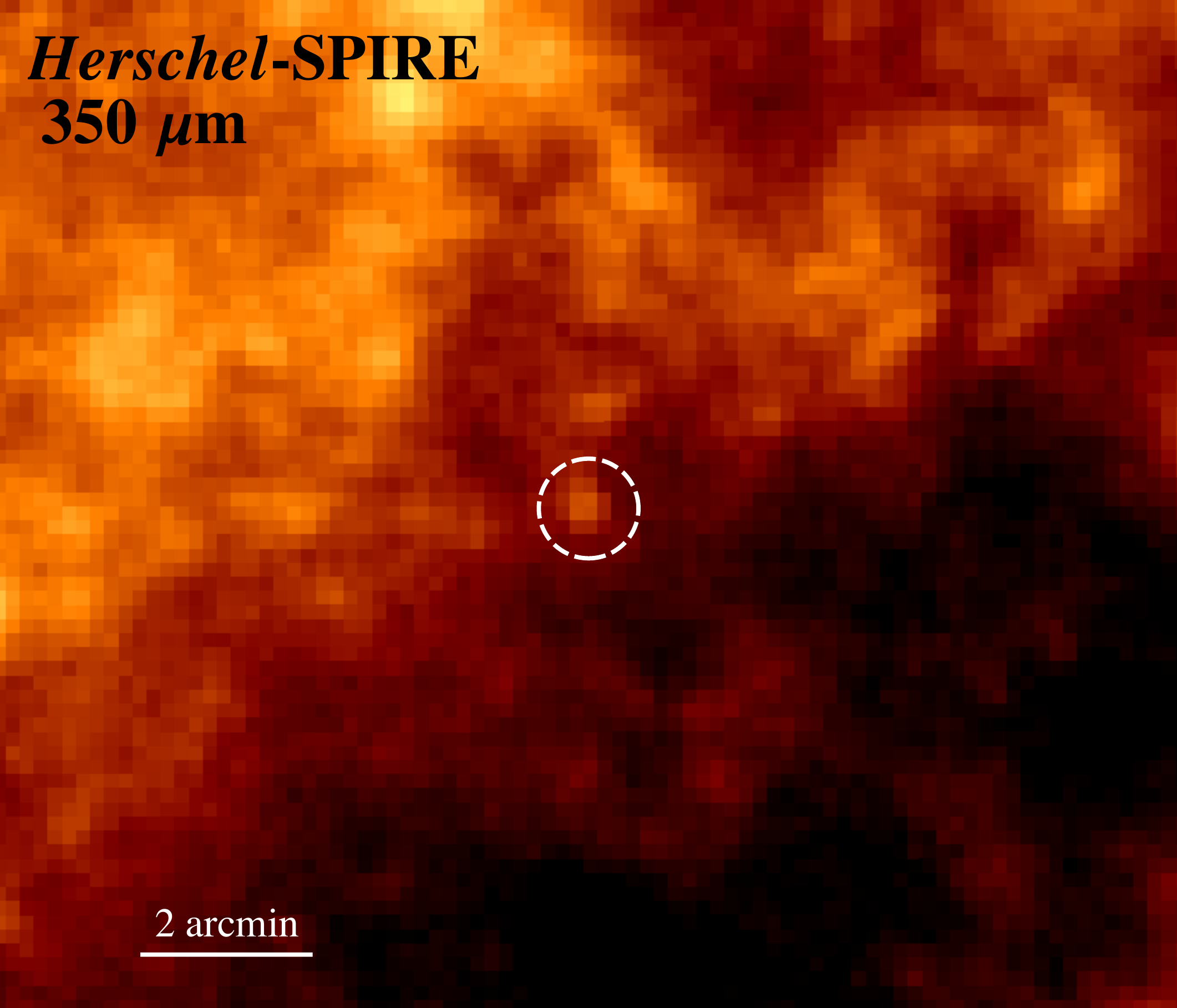}
\includegraphics[width=0.32\textwidth]{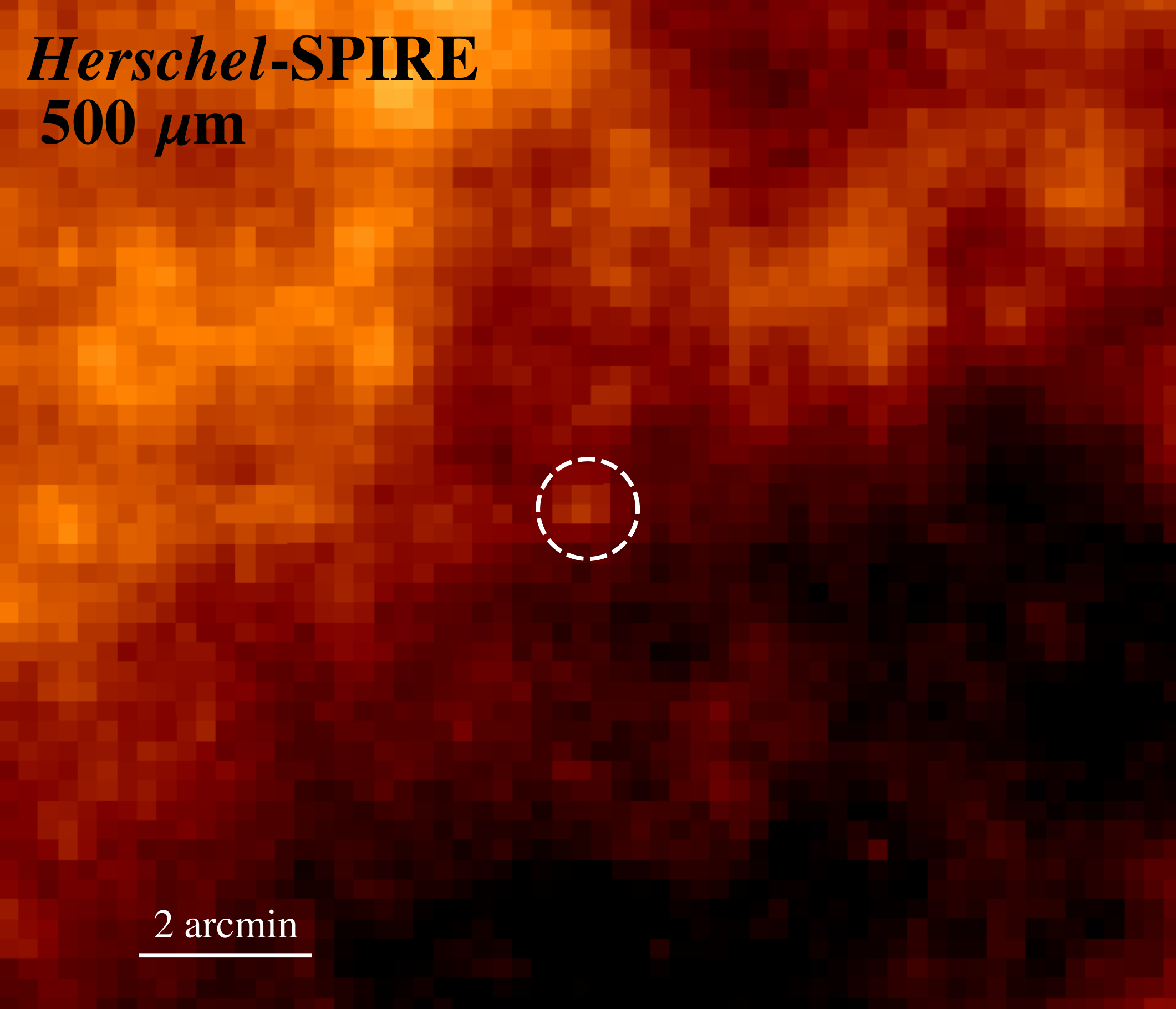}
\caption{Optical (H$\alpha$ - top left panel) and IR images of the WR nebula RCW\,58. The IR images were obtained by \textit{WISE} and \textit{Herschel} PACS and SPIRE. All panel have the same field of view. North
  is up and east to the left.}
\label{fig:rcw58mw}
\end{figure*}

In this paper, we analyse the nebular dust and gas properties of RCW\,58 around the WN8h star WR\,40 (HD\,96548). This WR nebula consists of a clumpy elliptical ring-like structure mostly detected in H$\alpha$ \citep[see Fig.~\ref{fig:rcw58mw} top left panel;][]{Stock2010} enclosed within faint, extended, smooth [O\,{\sc iii}] emission detected in narrow-band images. \citet{Gruendl2000} suggested that the offset between the H$\alpha$ and [O\,{\sc iii}] emission peaks is a result of radiative cooling behind a shock wave propagating into the circumstellar medium, where the outer [O\,{\sc iii}] {\it skin} traces the position of the expanding shock. The kinematics of RCW\,58 is poorly determined. Fabry-Perot spectra obtained by \citet{Chu1982} found chaotic motions consisting of random velocity variations on the order of 30~km~s$^{-1}$ and line-broadening or line-splitting of 60~km~s$^{-1}$.  High-resolution long-slit spectra taken at the bright south-east edge by \citet{Smith1988} are consistent with a shell expanding at $v_\mathrm{exp} = 87$~km~s$^{-1}$ containing clumps moving with velocities lower than the shell expansion velocity. Both shell and clumps have linewidths in excess of thermal broadening, indicative of turbulence.

\begin{table*}
\begin{center}
   \caption{Details of the IR observations used in this paper.}
   \begin{tabular}{lcccccc}
   \hline
   \hline
   Instrument                          & Date       & Obs. ID or Project         & PI                &    $\lambda_{\mathrm{c}}$   & Duration & Processing level \\
                                       &(yyyy-mm-dd)&                  &                   &    $(\mu$m)                 & (s)      & 
                    \\
   \hline
   \hline
   {\it WISE}                          & 2010-01-22 & 1677m652\_ac51    &                   &    12 (W3), 22 (W4)        & 8.8      &  Level 3         \\
   {\it Herschel} PACS                 & 2012-03-22 & 1342243127       & A.P.\,Marston&    70, 160                            & 1096     &  Level 3         \\
                                       &            & 1342243129       & A.P.\,Marston&    100, 160                            & 1096     &   Level 3         \\
   {\it Herschel} SPIRE                & 2012-03-03 & 1342241089       & A.P.\,Marston&    250, 350, 500                      & 1086      &  Level 2         \\
   \hline
  \end{tabular}
  \label{tab:obs}
\end{center}
\end{table*}

Optical spectroscopic studies of RCW\,58 have addressed the chemical abundances and physical properties of this WR nebula \citep[see, e.g.,][]{Kwitter1984,Rosa1990,Stock2011} and have confirmed strong nitrogen and helium enrichment. This is evidence that CNO-cycle nucleosynthesis products are present in the stellar ejecta material. The nebula is low-ionization and low-density. An electron temperature $T_\mathrm{e}=6450\pm220$~K and electron density $n_\mathrm{e}=100\pm100$~cm$^{-3}$ were derived from high-resolution spectra obtained recently by \citet{Esteban2016} using the Magellan Echellette at the 6.5~m Clay Telescope. The total mass of ionized gas in RCW\,58 has been estimated to be of order 3~M$_\odot$ from 11~cm radio flux measurements \citep{Smith1970}.

There are not many IR studies addressing the dust properties of RCW\,58. Its morphology is very similar to that of the optical image and the dust appears to be correlated with the H$\alpha$ clumps (see Fig.~\ref{fig:rcw58mw}). \citet{Marston1991} analysed {\it IRAS} observations at 50~$\mu$m and 100~$\mu$m of RCW\,58 along with other two WR nebulae (S\,308 and NGC\,6888). The main conclusions were that the dust in RCW\,58 has temperatures $\sim$33~K with a large variation (25--50~K) and that the estimated mass of dust is 0.39~M$_\odot$, implying a total nebula gas mass $\sim39$~M$_\odot$, far too big to have a stellar ejecta origin. As a result, \citet{Marston1991} suggested that RCW\,58 was composed mainly of swept up ISM. \citet{Mathis1992} used the same {\it IRAS} observations but analysed the 25~$\mu$m, 60~$\mu$m and 100~$\mu$m emission, and found that the IR properties of RCW\,58 can be explained by the presence of two populations of grains with size distributions 0.002--0.008~$\mu$m and 0.005-0.05~$\mu$m for the stellar radiation field and distance they assumed. The estimated mass of dust in their models is $<10^{-2}$~M$_{\odot}$, with a corresponding total nebular mass of $\sim$0.4~M$_{\odot}$. 
More recently, \citet{Toala2015} used {\it Wide-field Infrared Survey Explorer} ({\it WISE}) observations and showed that RCW\,58 is heavily contaminated by background ISM material mainly detected in the 12~$\mu$m band, which was not taken into account by these two previous IR studies.

The optical variability of WR\,40, the central star of RCW\,58, has been interpreted previously as originating from a WR$+$compact binary system \citep{Moffat1980}. However, \citet{Ramiaramanantsoa2019} presented recent {\it BRIght Target Explorer (BRITE)} constellation of nanosatellites observations of WR\,40 and demonstrated that the variability from this WN8 star is merely due to intrinsic stochastic variations in the wind. 

\citet{Ramiaramanantsoa2019} showed that starlight scattered by free electrons in a distribution of discrete clumps of different masses and lifetimes in WR\,40's wind can account for the variability in the light curve. If WR\,40 has a companion, as suggested for WN8 stars, it is well hidden within its the dense wind as suggested for WR\,124 \citep[see, e.g.,][]{Toala2018}.

This paper is organized as follows. In Section~\ref{sec:obs} we detail the IR and radio data used in our analysis as well as their preparation. Section~\ref{sec:work} describes the {\sc cloudy} models of RCW\,58 and in Section~\ref{sec:results} we present the model results. We discuss the grain size distributions, dust and nebular masses, and their implications for the evolution of RCW\,58 and its central star, WR\,40 in Section~\ref{sec:disc}, and summarize our conclusions in Section~\ref{sec:summ}.
 
\section{OBSERVATIONS}
\label{sec:obs}

\subsection{Data}
\label{subsec:data}

We use public IR observations of RCW\,58 from the \textit{WISE} and \textit{Herschel} space telescopes that together 
cover the 12--500~$\mu$m wavelength range. All the observations were retrieved 
from the NASA/IPAC Infrared Science Archive\footnote{\url{https://irsa.ipac.caltech.edu/frontpage/}} and their details, such as date, observation ID, PI and processing level, are listed in Table~\ref{tab:obs}.

Figure~\ref{fig:rcw58mw} shows all of the IR images obtained from the \textit{WISE} telescope, and the \textit{Herschel} PACS and SPIRE instruments, together with an H$\alpha$ image retrieved from the SuperCOSMOS Sky Survey\footnote{\url{http://www-wfau.roe.ac.uk/sss/halpha/}} \citep{Parker2005}. RCW\,58 is clearly detected in most images except in the \textit{Herschel} SPIRE images, which are dominated by background cold interstellar material. We note that the \textit{Herschel} PACS images do not entirely cover RCW\,58. The observations cover the NW, NE, SE and SW regions of the nebula (see Figure~\ref{fig:rcw58mw}).

\begin{figure}
\centering
\includegraphics[width=0.5\textwidth]{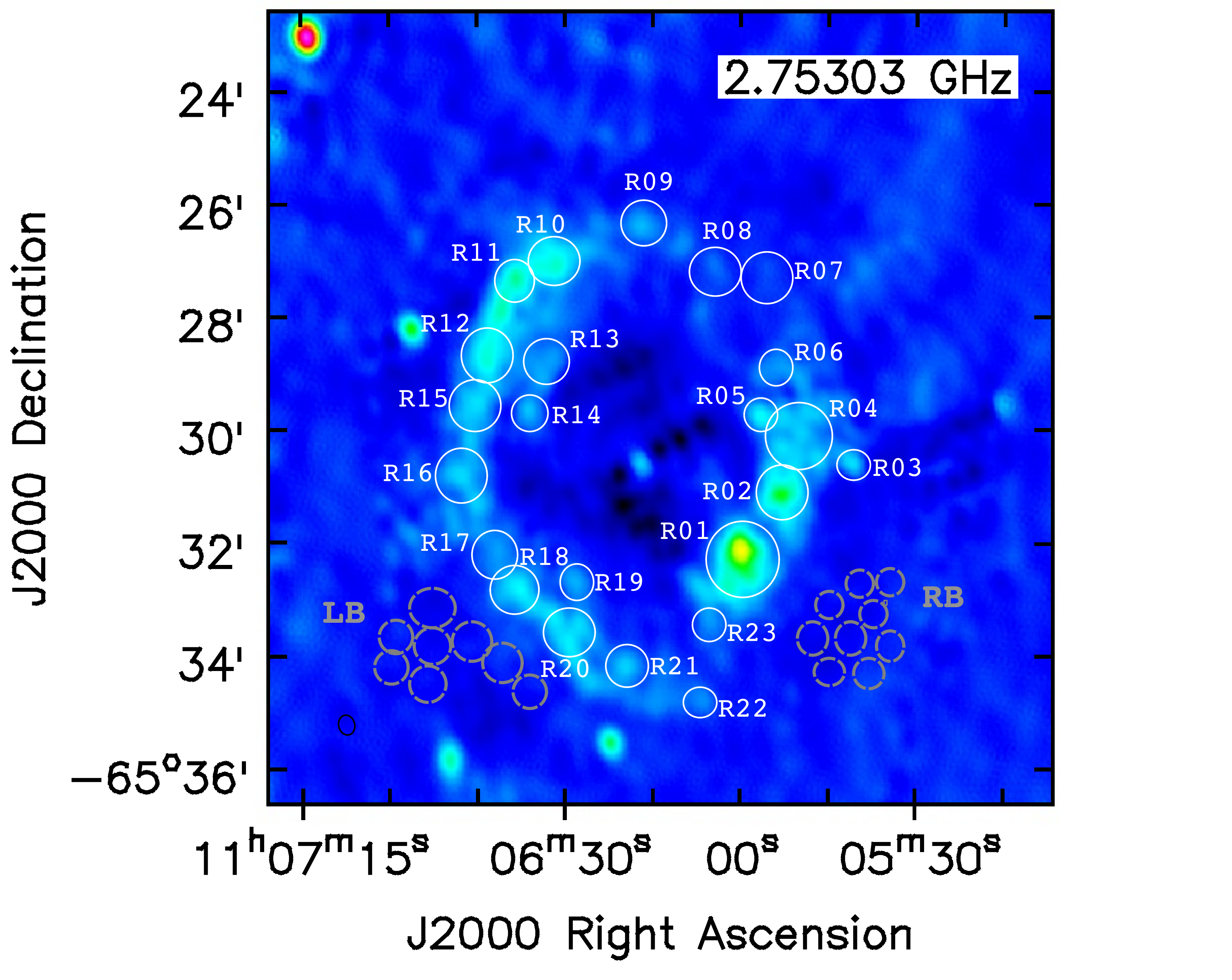}
\caption{ATCA image of 2.75 GHz continuum in Jy/beam units of RCW\,58. The synthesized beam FWHM is 21.3 $\times$ 17.1~arcsec and the rms noise between 0.2 to 0.89~mJy beam$^{-1}$. White Regions were used to get a density estimation. RB and LB gray regions were used to estimate the background correction in the C1 and C2 photometry process, respectively.}
\label{fig:rcw58atca}
\end{figure}

Additionally, we complemented our IR images with radio observations obtained with the Australian Telescope Compact Array (ATCA) to extend the wavelength range of our study. The ATCA observations have a total exposure time of 44.53 hr in the 750A configuration and were taken on 2010 June 30 under project C2618 (PI: J.A.\,Toal\'{a}). We created a 10 cm continuum image with a central frequency $\nu_{0} = 2.735$ GHz (10.9 cm), an rms noise from 0.2 to 0.89~mJy beam$^{-1}$, and a synthesized beam with FWHM = 21.3 $\times$ 17.1~arcsec.

\subsection{Photometry}
\label{subsec:phot}

The photometry extraction procedure used in this work is essentially the same as that described in Paper I \citep[see also][]{Rubio2020}. To start, we defined an extraction region that included nebular emission with a background contribution. To obtain just the nebular emission we estimated the background contribution and its variation (the measurement error) and took into account instrumental effects (the calibration error). The background variation was calculated by selecting several different regions from the vicinity of the nebula. The total error ($\sigma_\mathrm{Tot}$) was then computed using
\begin{equation}
    \sigma_\mathrm{Tot} = \sqrt{\sigma_\mathrm{Back}^2 + \sigma_\mathrm{Cal}^2}\ ,
\end{equation}
\noindent where $\sigma_\mathrm{Back}$ is the measurement error and $\sigma_\mathrm{Cal}$ is the calibration uncertainty.  The calibration uncertainty is a percentage of the flux measurement for each instrument. In particular, the \textit{WISE} error is 4.5 and 5.7 per cent from the W3 and W4 bands, respectively\footnote{ See the Explanatory Supplement to the WISE Preliminary Data Release Products at \url{https://wise2.ipac.caltech.edu/docs/release/prelim/expsup/}.}.

\begin{figure*}
\centering
\includegraphics[width=0.32\textwidth]{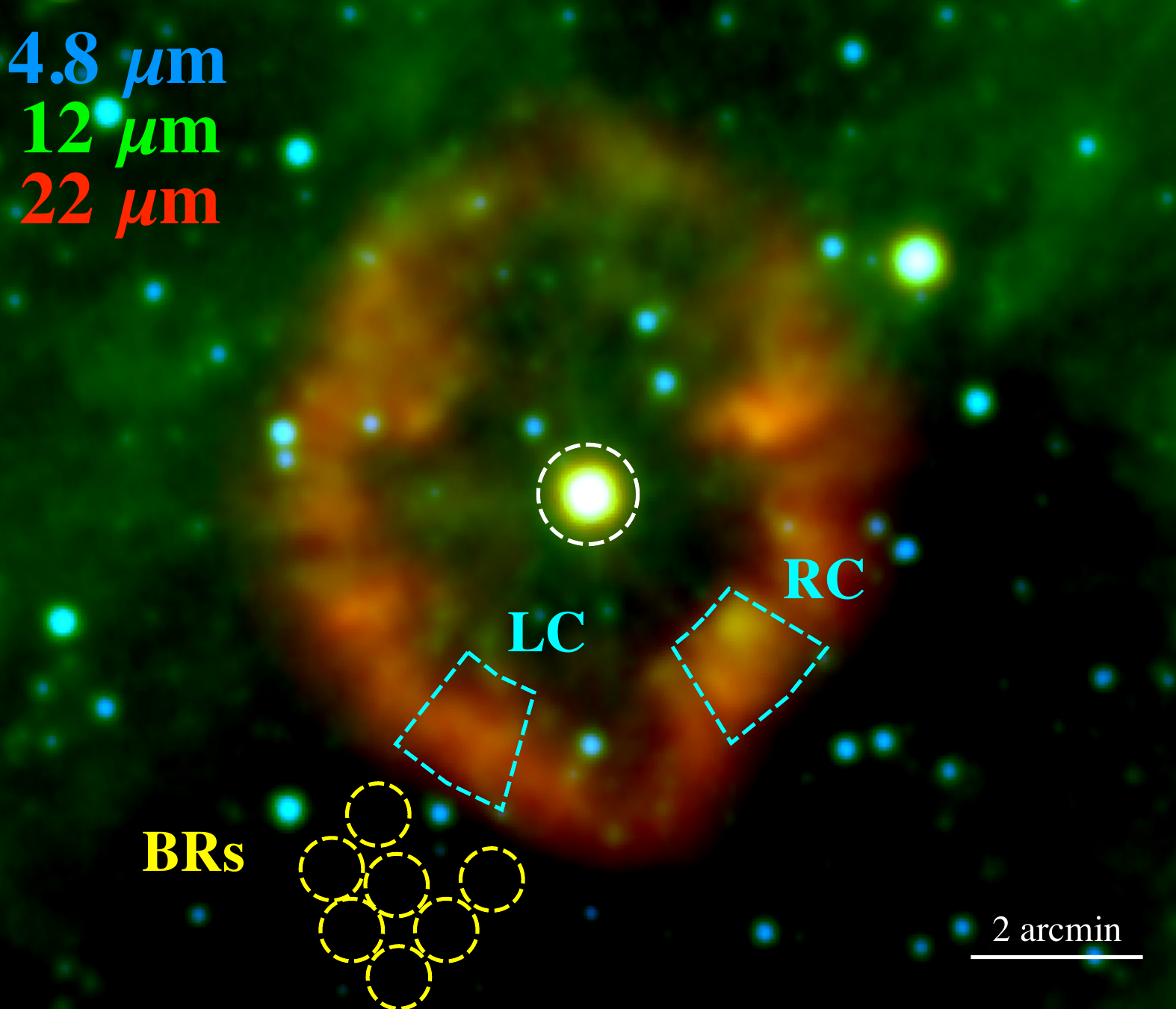}
\includegraphics[width=0.32\textwidth]{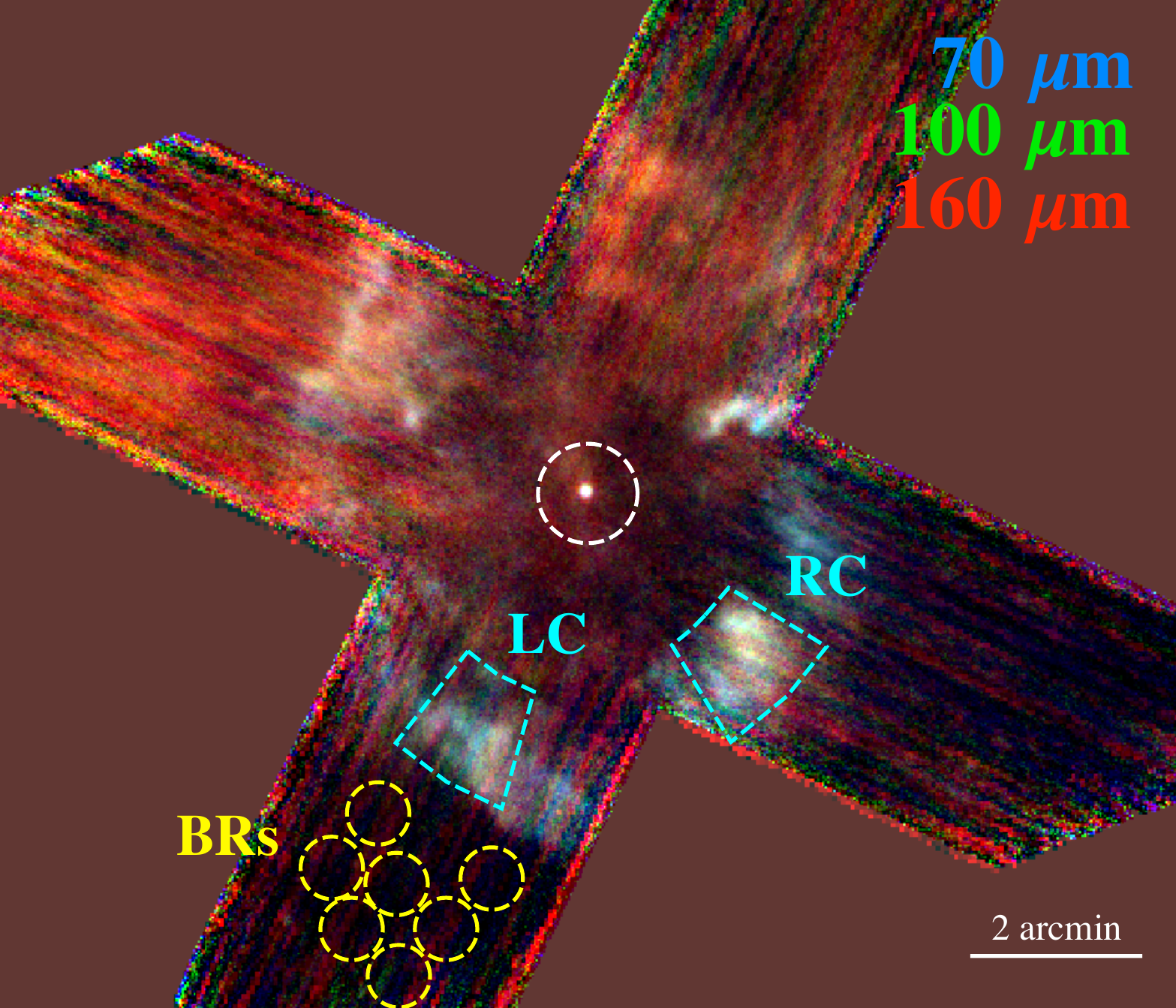}
\includegraphics[width=0.32\textwidth]{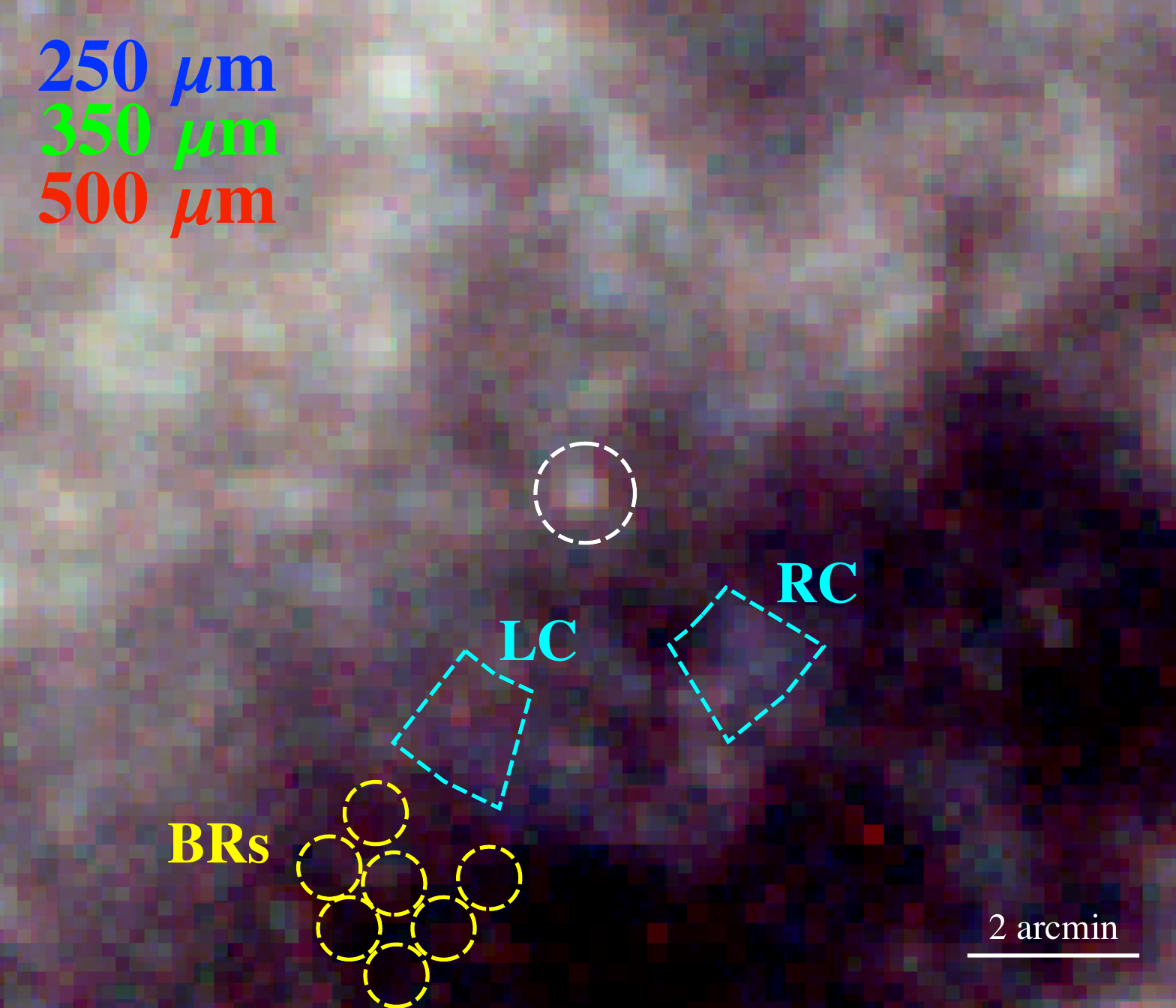}
\caption{Color-composite IR images of RCW\,58. The left, middle and right panels correspond to \textit{WISE},\textit{Herschel} PACS and SPIRE observations, respectively. The selected regions for study in RCW\,58 are indicated with 
cyan polygonal shapes (RC and LC). Background regions are represented by (yellow) dashed-line circular apertures. The position of the central star, WR\,40, is indicated in the centre of each panel by a (white) circular region. All images have same the Field of View (FoV). North
  is up and East is to the left.}
\label{fig:rcw58rs}
\end{figure*}

We attempted to obtain a global spectral energy distribution (SED) of RCW\,58 by selecting an elliptical region encompassing the nebular emission delimited in the H$\alpha$ image (see Figure~\ref{fig:rcw58mw} upper left panel). However, at IR wavelengths there is a strong gradient of emission from the north to the south of the image panels in Figure~\ref{fig:rcw58mw}, which we attribute to background ISM since it does not appear to attenuate the optical image. Furthermore, the incomplete coverage of the \textit{Herschel} PACS observations severely restricted our choices for background regions to use in the extraction process. As a result, it proved impossible to produce a satisfactory global SED. This precluded an appropriate analysis of the global properties of RCW\,58. Consequently, we decided to restrict our analysis to two specific clumps in the southern part of RCW\,58, situated in regions relatively free from background contamination.

We selected two bright clumps located in the nebular ring towards the southwest (SW) and southeast (SE) of the central star WR\,40. These clumps are labeled RC and LC and are illustrated in Figure~\ref{fig:rcw58rs}. This figure also shows the regions used for the background extraction regions (labeled BRs). These are located external to the nebula towards the SE direction. 
We note that the background is not homogeneous and to alleviate this issue we created seven independent regions which were used to generate an average background. This situation is similar when extracting fluxes from the ATCA image. The corresponding background regions for the ATCA photometry are shown in Figure~\ref{fig:rcw58atca} with gray dashed-line circular apertures.

\begin{figure*}
\centering
\includegraphics[width=1.0\textwidth]{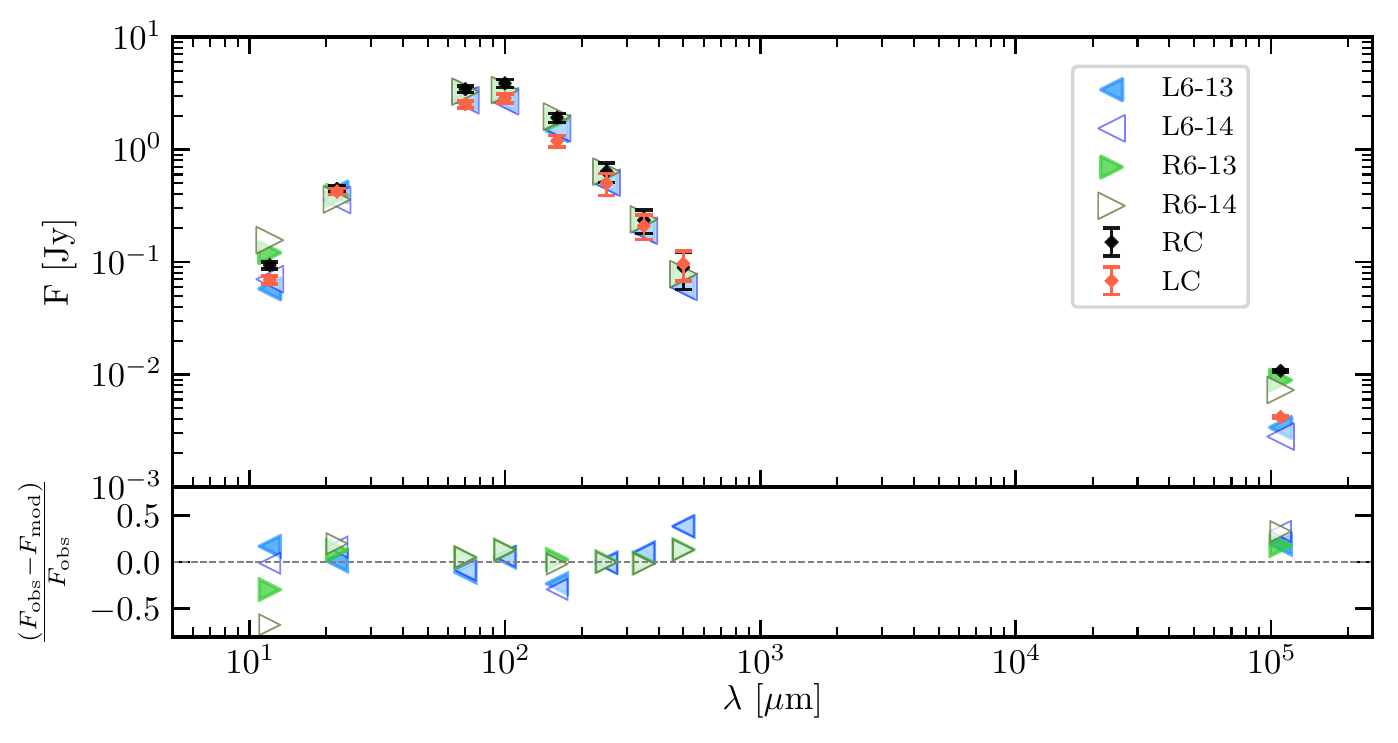}
\caption{SED obtained from the IR and ATCA observations of RCW\,58: black and red diamonds correspond to the RC and LC regions, respectively. The error bars were obtained from the instrument calibration and background 
subtraction process. The synthetic SED data points, obtained from our best models of RC (R\,6-13 and R\,6-14) and LC (L\,6-13 and L\,6-14), are represented by triangles.} 
\label{fig:seds}
\end{figure*}

In Table~\ref{tab:fot}, we present the flux estimates for each IR and radio measurement of RC and LC with their corresponding uncertainties ($\sigma_\mathrm{Tot}$). The final IR SED of RC and LC regions are presented in Figure~\ref{fig:seds}. The SEDs for both clumps are very similar, having broad peaks with maxima between $70~\mu$m and $100~\mu$m.

\begin{table}
  \begin{center}
     \caption{Flux densities and uncertainties of different IR band
       observations of the selected RC and LC regions in RCW\,58. {\bf $\sigma_\mathrm{Back}$}  is the
       uncertainty associated with background inhomogeneity,
       $\sigma_{\mathrm{Cal}}$ is the uncertainty associated with the
       instrument calibration of each observation and
       $\sigma_{\mathrm{Tot}}$ is the sum of these uncertainties.}
     \begin{tabular}{cccccc}
     \hline
     \hline
     Instrument  &$\lambda_\mathrm{c}$ & Flux              & {\bf $\sigma_\mathrm{Back}$} & $\sigma_{\mathrm{Cal}}$ & $\sigma_{\mathrm{Tot}}$ \\
                          &$[\mu$m]  & [Jy]              &  [Jy]                     &  [Jy]                        & [Jy]                        \\
     \hline
     RC region\\
     \hline
     {\it WISE}            & 12        & 0.0936  &  0.0054  & 0.0042 &  0.0068    \\
                           & 22        & 0.4471  &  0.0130  & 0.0255 &  0.0286  \\
     {\it Herschel} PACS   & 70        & 3.4458  &  0.1508  & 0.1723 &  0.2290    \\
                           & 100       & 3.8710  &  0.2496  & 0.1936 &  0.3158   \\
                           & 160       & 1.9205  &  0.1507  & 0.0960 &  0.1787   \\
     {\it Herschel} SPIRE  & 250       & 0.6344  &  0.1167  & 0.0444 &  0.1249   \\
                           & 350       & 0.2351  &  0.0531  & 0.0165 &  0.0556   \\
                           & 500       & 0.0897  &  0.0319  & 0.0063 &  0.0325   \\
                           &&&&&\\
                         &$[\mu$m]  & [mJy]              &  [mJy]                     &  [mJy]                        & [mJy]                        \\
                           &&&&&\\
    ATCA                   & 1.09$\times10^{5}$    & 10.8 &  0.295 &  -    &  0.295\\
    \hline                           
    LC region\\
     \hline
     {\it WISE}            & 12        &  0.0694 &  0.0046 &  0.0031 & 0.0055     \\
                           & 22        &  0.4234 &  0.0111 &  0.0241 & 0.0266     \\
     {\it Herschel} PACS   & 70        &  2.5268 &  0.1292 &  0.1263 & 0.1807     \\
                           & 100       &  2.8545 &  0.2137 &  0.1427 & 0.2570    \\
                           & 160       &  1.1943 &  0.1289 &  0.0597 & 0.1420    \\
     {\it Herschel} SPIRE  & 250       &  0.5012 &  0.1006 &  0.0351 & 0.1122    \\
                           & 350       &  0.2098 &  0.0468 &  0.0147 & 0.0512    \\
                           & 500       &  0.0968 &  0.0274 &  0.0068 & 0.0291    \\
                           &&&&&\\
                           &$[\mu$m]  & [mJy]              &  [mJy]                     &  [mJy]                        & [mJy]                        \\
                           &&&&&\\
    ATCA                   & 1.09$\times10^{5}$  &  4.18   & 0.106 &  -    & 0.106    \\
     \hline
     \hline
     \end{tabular}
     \label{tab:fot}
  \end{center}
\end{table}

At mid-infrared wavelengths, $\lambda \ge 100~\mu$m, optically thin thermal dust emission at a mean temperature $T_\mathrm{d}$ can be expressed as a modified black-body (MBB) with a power law:

\begin{equation}
    F_\nu = M_\mathrm{d}~\kappa_{\nu 0} \left( \frac{\nu}{\nu_0}\right)^{\beta} \frac{B_\nu(T_\mathrm{d})}{d^2},
\end{equation}
where $d$ is the distance to the nebula, $B_\nu(T_\mathrm{d})$ is the Planck function, $M_\mathrm{d}$ is the total mass of dust, $\kappa_{\nu 0}= 1.92$~cm$^2$g$^{-1}$ is the dust emissivity normalization at the wavelength $\nu_0=350~\mu$m \citep{Draine2003}. The emissivity index $\beta$ depends on the dust properties, for example, $\beta=2$ is representative of crystalline grains, while $\beta=1.2$ is consistent with amorphous carbon dust \citep{Zubko1996}. An iterative fitting procedure (see Paper I) returns the mean temperature $T_\mathrm{d}$, the power-law index $\beta$ and the dust mass $M_\mathrm{d}$ responsible for the long wavelength ($\lambda \ge 100~\mu$m) emission.

For RC in RCW\,58 we obtained $M_\mathrm{d}=0.029 \pm0.006$~M$_{\odot}$, $T_\mathrm{d}=44.41 \pm2.10$~K and $\beta=1.43\pm0.10$, while for LC  we found $M_\mathrm{d}=0.041 \pm0.007$~M$_{\odot}$, while dust temperature is  $T_\mathrm{d}=44.18 \pm2.22$~K and $\beta=1.5\pm0.1$. The similarity in the mean temperature and emissivity index of the two clumps suggests that they share general dust properties.

We note that the SED peak is too broad to be fit by a single population of grains at a uniform temperature, and detailed modeling will therefore be required to give the full picture of the grain populations and their spatial distributions.

\subsection{Free-free emission}
\label{ssec:ffe}

Radio continuum observations have been used in the past to study the distribution of the ionized gas, the electron density and the ionized mass in a sample of WR ring nebulae \citep{Cappa1999,Cappa2006}. 
Our previously unpublished, high angular resolution ATCA observation, presented in Figure~\ref{fig:rcw58atca}, displays the elliptical clumpy morphology of RCW\,58 and corresponds very closely to the optical image shown in Figure~\ref{fig:rcw58mw}.

The ATCA observation provides an additional photometric measurement, which extends our IR SED to longer wavelengths. We used extraction regions similar to RC and LC described above (see Figure~\ref{fig:rcw58rs}). The 2.75~GHz flux measurements of the RC and LC regions are listed in Table~\ref{tab:fot} and are also plotted alongside the IR photometry in Figure~\ref{fig:seds}, where the corresponding wavelength is $\lambda = 1.09\times10^5 \mu$m.

We used the ATCA observations to estimate electron densities, $n_\mathrm{e}$, in 23 different clumps identified in the main elliptical shell of RCW\,58 (see Figure~\ref{fig:rcw58atca}). First, we calculated the brightness temperature, $T_\mathrm{B}$, for each clump by assuming that the radio emission in RCW\,58 is due to free-free radiation and is optically thin. $T_\mathrm{B}$ is defined as \citep{Wilson2009}
\begin{equation}
T_\mathrm{B} = \left( \frac{c^2}{2 k_\mathrm{B} \nu_0^2}\right) \left(\frac{S}{\Omega}\right) \ ,
\label{eqn:tbright}
\end{equation}
\noindent where $c$ is the speed of light, $k_\mathrm{B}$ is the Boltzmann constant, $S$ is the measured flux at frequency $\nu_0$ and $\Omega$ is the angular size of each clump. Next, the emission measure, EM, can be expressed as
\begin{equation}
\mathrm{EM} = 12.143\, \tau \, \left(\frac{\nu_0}{[\mathrm{GHz}]}\right)^{2.1} \left(\frac{T_\mathrm{e}}{[\mathrm{K}]}\right)^{1.35},
\label{eqn:EM}
\end{equation}
\noindent where $\tau = T_\mathrm{B}/T_\mathrm{e}$ and $T_\mathrm{e}$ is the electron temperature. Finally, the electron density, $n_\mathrm{e}$, can be expressed as
\begin{equation}
\label{eqn:ne}
  n_\mathrm{e} = \sqrt{\mathrm{EM}/ D} \ ,
\end{equation}
\noindent where $D$ is the depth of the nebular material in pc along the line of sight. We assumed that the line-of-sight depth of each clump is the same as its diameter, $D$.

\begin{figure}
\centering
\includegraphics[width=0.5\textwidth]{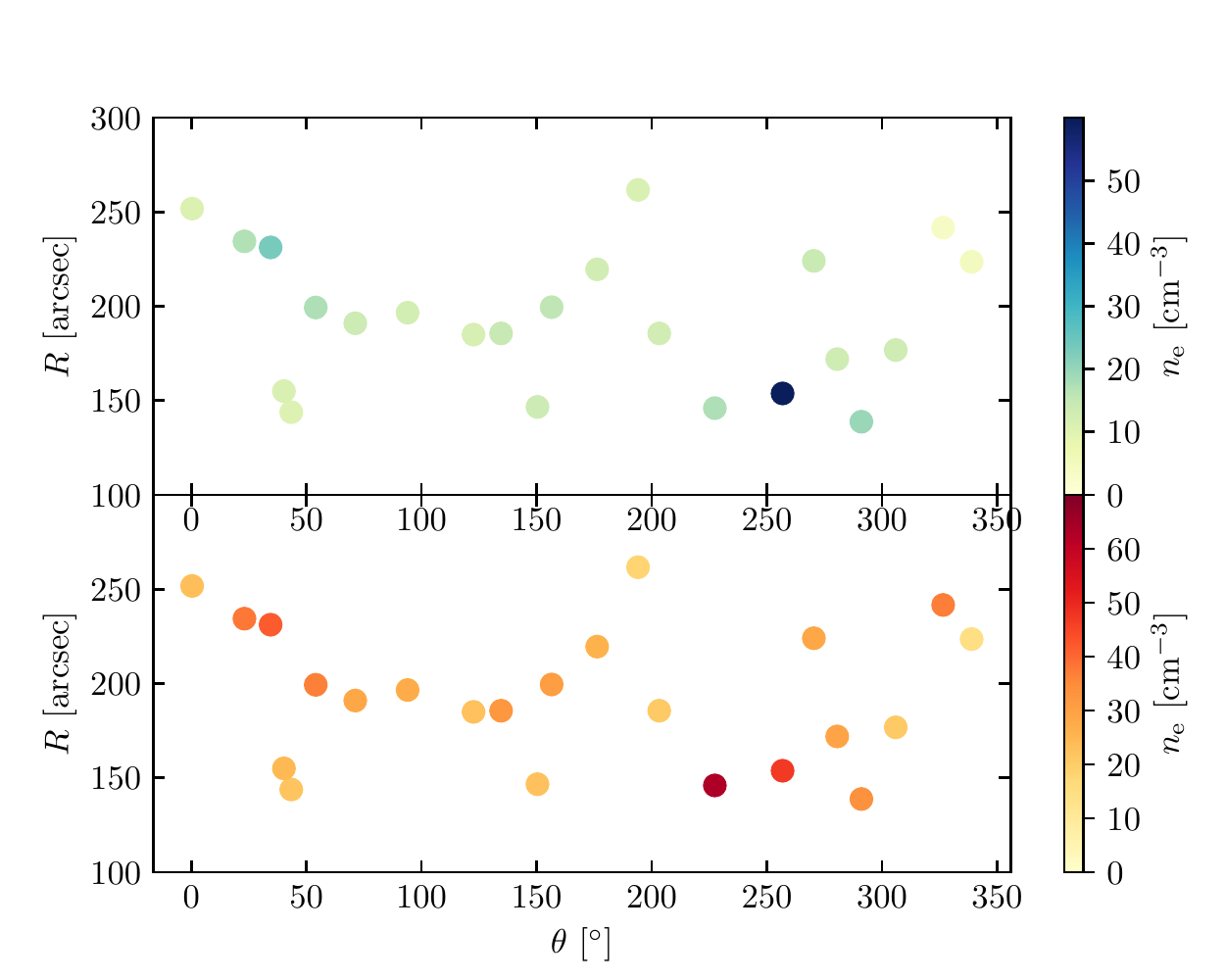}
\caption{Electron density distribution as a function position angle and distance with respect to the central star ($\theta,R$). Upper panel shows the peak densities, and bottom panel shows average densities. Each point corresponds to a single region of Figure~\ref{fig:rcw58atca}.}
\label{fig:den}
\end{figure}

The resultant $n_\mathrm{e}$ values obtained for each clump represent the average electron densities and are illustrated in the lower panel of  Figure~\ref{fig:den} as a function of position angle and distance from the central star. We also estimated the peak $n_\mathrm{e}$ for each circular region by adopting the major axis of the beam as the diameter (and depth) of each clump. The peak $n_\mathrm{e}$ values are shown in the upper panel of Figure~\ref{fig:den}.

From Figure~\ref{fig:den} we see that the range of values for both the average and peak electron densities is between 10 and 60~cm$^{-3}$. The highest values are found for position angles $\theta = 45^\circ \pm 15^\circ$ (regions R10, R11 and R12) and $\theta = 245^\circ \pm 15^\circ$ (regions R01 and R02), which are almost exactly diametrically opposed.
Clump R01 is contained within the RC region described in \S~\ref{subsec:phot}. For comparison, we note that 
an upper limit of $<100$~cm$^{-3}$ was found by \citet{Esteban2016} using the [O\,{\sc ii}] 3726,3729~\AA\, doublet.

\section{Dust modeling of RCW\,58}
\label{sec:work}

As noted in the previous sections, a global model of the dust and nebular properties of RCW\,58 is hampered by the extended background emission in the northern part of the nebula. For this reason, regions RC and LC  (see \S~\ref{subsec:phot}) have been selected as representative of the properties of the ring nebula. Whilst LC does not appear to be special at any wavelength, RC harbours the region with highest average electron density as derived from the ATCA observations.

Following our previous work in Paper I, we will use the spectral synthesis code {\sc cloudy} \citep[version 17;][]{Ferland2017} to model the gas and dust properties of these two regions in RCW\,58. In addition, {\sc pycloudy} \citep{Morisset2006} will be used to produce synthetic optical and infrared observations, specifically, the spectra and IR photometry which can be directly compared to the available observations.

{\sc cloudy} takes as input parameters the stellar atmosphere (radiation field and luminosity), the gas density distribution and chemical abundances, and the dust properties (grain size distribution and composition). Both gas and dust must be included in the photoionization models since each component absorbs and reprocesses a fraction of the incident stellar EUV radiation. {\sc cloudy} solves the radiative transfer in 1D spherical geometry, while {\sc pycloudy} is used to generate user-defined 3D nebulae from the {\sc cloudy} output. In the following subsections, we detail the parameters used in our model.

\subsection{Stellar atmosphere}
\label{ssec:stell}

WR\,40 is a WN8h spectral type-star \citep{Smith1996}. Detailed modelling of the stellar spectrum over a wide wavelength range using line-blanketed, non-LTE model atmospheres and taking into account clumping in the expanding stellar wind has been performed by \citet{Herald2001} and \citet{Hamann2006}.

The models are characterized by the stellar temperature $T_\star$, the chemical abundances and the transformed radius  
\begin{equation}
  R_\mathrm{t} = R_\star \left[  \frac{v_\infty}{2500~\mathrm{km\,s}^{-1}} \bigg/ \frac{\dot{M}\sqrt{D_\infty}}{10^{-4} \mathrm{M}_\odot~\mathrm{yr}^{-1}}  \right]^{2/3}
    \label{eq:trlaw}
\end{equation}
where $R_\star$ and $T_\star$ are the stellar radius and temperature defined at a Rosseland optical depth of 20, $v_\infty$ and $\dot{M}$ are the stellar wind terminal velocity and (clumped) mass-loss rate, and $D_\infty$ is a clumping factor. Models with the same stellar temperature and transformed radius produce resultant spectra having the same equivalent widths \citep{Schmutz1989}. This scaling relation means that different sets of values of $(R_\star,v_\infty,\dot{M},D_\infty)$ can give the same stellar spectrum. 

\citet{Herald2001} found that their best model was achieved with values $(10.6~\mathrm{R}_\odot,840\,\mbox{km\,s$^{-1}$},3.2\times 10^{-5}~ \mathrm{M}_\odot\mbox{yr$^{-1}$},10)$ together with $T_\star = 45.4$~kK, where $v_\infty$ was adopted from \citet{1995CrowtherSmithHillierSchmutz}. The corresponding value of the transformed radius is $R_\mathrm{t} = 5.16~\mathrm{R}_\odot$. On the other hand, 
 \citet{Hamann2006,Hamann2019} using the stellar atmosphere code {\sc powr}\footnote{\url{http://www.astro.physik.uni-potsdam.de/~wrh/PoWR/powrgrid1.php}},
find a good model fit to WR\,40 corresponding to the set of values $(14.5~\mathrm{R}_\odot,650\,\mbox{km\,s$^{-1}$},6.3\times10^{-5}~\mathrm{M}_\odot\mbox{yr$^{-1}$},4)$ together with $T_\star=44.7$~kK, which take into account the corrections for the new \textit{Gaia} distances \citep{Rate2020}.
For this set of values, the transformed radius is $R_\mathrm{t} = 5.27~ \mathrm{R}_\odot$. Hence the two quite different sets of numbers lead to very similar model parameters. \citet{Hamann2006,Hamann2019}\ recommend using their model labelled 06-14 from the WNL-H20 grid with Galactic metallicity, which corresponds to $\log R_\mathrm{t} = 0.7$, $T_\star = 44.7$~kK and a surface hydrogen abundance of 20 per cent by mass. Details of this model and the grid-adjacent model 06-13, which has $\log R_\mathrm{t} = 0.8$, are listed in Table~\ref{tab:powr}. We note that the main difference in increasing $R_\mathrm{t}$ is the higher rate of helium-ionizing photons.

The stellar luminosity, temperature and radius are related through the Stefan-Boltzmann law. The luminosity of WR\,40 can be estimated using the $v$ magnitude, the reddening law $R_V$, $E_{b-v}$ and an assumption as to either the distance $d$ or the absolute magnitude $M_v$. Using the most recent distance of 3.83~kpc \citep{Rate2020}, we have rescaled both the  \citet{Herald2001} luminosity and the \citet{Hamann2019} luminosity and in each case we obtain $\log(L/\mathrm{L}_{\odot})=5.9$, so this is the value we adopt in our models. The {\sc powr} grid of models is normalized to $\log(L/\mathrm{L}_\odot)=5.3$, so the hydrogen and helium ionizing photon rates listed in Table~\ref{tab:powr} should be scaled accordingly.

\begin{figure*}
\centering
\includegraphics[width=0.9\textwidth]{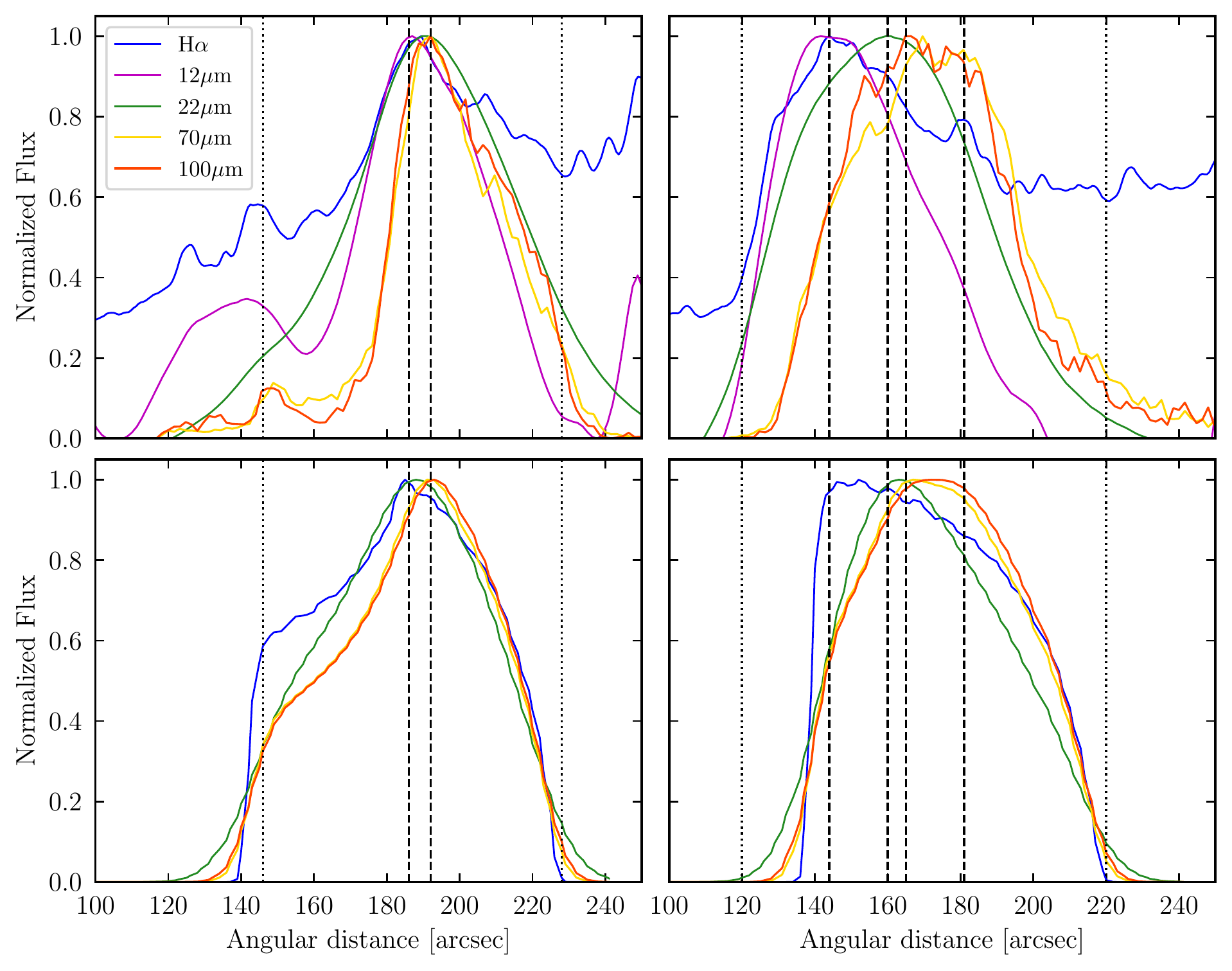}
\caption{{\it Top panels}. Surface brightness profile as a function of angular distance from the star
integrated over LC (PA: $196^{\circ}\leq \theta \leq 218^{\circ}$ - left) and RC (PA: $122^{\circ}\leq \theta \leq 150^{\circ}$ - right) clumps. Different colours represent profiles obtained from different instruments. The dashed vertical lines show the position of the emission peaks whilst dotted lines denote the inner and outer radii of the RC and LC regions. {\it Bottom panels}. Simulated surface brightness profiles obtained from our best models of the LC (left) and RC (right). The colour code is the same as for the top panels.}
\label{fig:profiles}
\end{figure*}

\begin{table}
  \begin{center}
     \caption{Parameters of {\sc powr} Models WNL\,06-13 and WNL\,06-14. WNL\,06-13 is adopted in this work to be the stellar atmosphere in our models, while WNL\,06-14 was found by \citet{Hamann2006} to be the best fit
       to the stellar spectrum of WR\,40.}
     \begin{tabular}{lcc}
     \hline
     \hline
     Model                                      & WNL\,06-13  &  WNL\,06-14           \\
     \hline
     $X_{\mathrm{H}}$ [\%]                      &    20       & 20                  \\
     $T_\star$ [kK]                      &    44.7     & 44.7                \\
     $\log_{10}R_{\mathrm{t}}$  [R$_{\odot}$]   &    0.8      & 0.7                 \\
     $\log_{10}Q_{\mathrm{H}}$  [s$^{-1}$]      &    48.95    & 48.85               \\
     $\log_{10}Q_{\mathrm{He}}$ [s$^{-1}$]      &    46.99    & 44.91               \\
     $D_{\infty}$                               &    4        &  4                  \\
     \hline
     \end{tabular}
     \label{tab:powr}
  \end{center}
\end{table}

\subsection{Nebular gas properties}
\label{ssec:nprop}

RCW\,58 consists of slow-moving clumps of photoionized stellar ejecta enveloped by a shell of low-density material swept-up by the wind-blown bubble of the central star \citep{Chu1982,Smith1988,Gruendl2000}. The shell is detected as faint [O\,{\sc iii}] emission extending beyond the H$\alpha$ clumps and corresponds to the outer shock of the wind bubble. 
 
The photoionized clumps form a roughly elliptical ring (see top-left panel of Fig.~\ref{fig:rcw58mw}) and often have sharp inner edges and diffuse outer edges \citep{Chu1982}. 

Electron densities have been estimated using the [S\,{\sc ii}] $\lambda\lambda$\,6717,6731 doublet and values between 200--500~cm$^{-3}$ were reported for positions corresponding to our region LC \citep{Smith1988}, although the uncertainties are of the same order since the emission lines are very faint. More recently, \citet{Esteban2016} derived $n_\mathrm{e}= 100\pm100$~cm$^{-3}$ from the 
[S\,{\sc ii}] $\lambda\lambda$\,6717,6731 doublet
observed using the 6.5~m Magellan Clay Telescope. For these observations, the $10\times 1$~arcsec$^{2}$ slit was positioned in the brightest region of the RC clump. \citet{Stock2011} also derived electron densities close to the low-density limit using the [S\,{\sc ii}] $\lambda\lambda$\,6717,6731 doublet from spectroscopic observations across the southwest of RCW\,58. Given the large uncertainties, these reported electron densities are  entirely consistent with our values determined from the ATCA radio observations (see \S~\ref{ssec:ffe}).

The chemical abundances in the nebula were studied by 
\citet{Kwitter1984}, \citet{Rosa1990} and \citet{Esteban2016}. \citet{Rosa1990} remarked that RCW\,58 is such a low ionization nebula that it is difficult to reliably determine $\mathrm{He}/\mathrm{H}$ because increased He abundance can mimic a lower stellar effective temperature since all the available helium-ionizing photons are absorbed in the inner part of the nebula. Note that for the stellar atmosphere models described above (see \S~\ref{ssec:stell}), the stellar effective temperature, defined at an optical depth of $\tau=2/3$, is much lower than $T_*$; in particular, $T_\mathrm{eff}\sim35$~kK for these models. All of the cited works find evidence for strong nitrogen and helium enrichment in RCW\,58, although the exact helium abundance is very uncertain. Such enrichment is an indication that CNO-cycle nucleosynthesis products are present in the stellar ejecta material. In  Table~\ref{tab:abun} we list the chemical abundances reported by \citet{Esteban2016} and \citet{Mendez2020}, which are the ones we adopt for our {\sc cloudy} models.

\begin{table}
  \begin{center}
  \caption[]{Chemical abundances in RCW\,58 used in our {\sc cloudy} models.}
     \begin{tabular}{ccl}
     \hline
    Element  & 12+log$_{10}$(X/H)  & Reference \\
     \hline
     He      &    11.23$\pm$0.02    &    \citet{Mendez2020} \\
     O       &    8.60$\pm$0.16     &    \citet{Mendez2020} \\
     N       &    8.67$\pm$0.18     &    \citet{Mendez2020} \\
     S       &    7.02$\pm$0.12     &    \citet{Esteban2016} \\
     Ar      &    6.44$\pm$0.24     &    \citet{Esteban2016} \\
     Fe      &    6.87$\pm$0.38     &    \citet{Esteban2016} \\
     \hline
     \end{tabular}
     \label{tab:abun}
  \end{center}
\end{table}

Finally, the first direct determination of the electron temperature in RCW\,58 was made by \citet{Esteban2016}, who obtained the value  $T_\mathrm{e}=6450\pm220$~K using the [N\,{\sc ii}] $5755/(6548+6584)$ emission-line diagnostic ratio.

\subsection{Dust properties}
\label{ssec:dprop}

Post-main-sequence massive stars produce silicate dust because their envelopes are oxygen rich owing to nucleosynthesis by the CNO cycle in the stellar cores \citep[e.g., ][]{Gail2005,Cherchneff2013}. The condensates that coalesce to form dust grains will include metal oxides and  calcium-magnesium-iron silicates. However, the dust production process in massive stars is still poorly understood and is highly dependent on the particular physical and chemical conditions in the region where it forms.

{\sc cloudy} provides optical constants for carbonaceous grains (graphite, amorphous carbon and long and short PAH molecules), silicon carbide, and astronomical amorphous silicate, namely olivine (MgFeSiO$_{4}$) \citep[e.g.,][]{Draine2003}. Only the last of these is of interest for RCW\,58 since the dust will have formed in an oxygen-rich environment.

The dust mixture in RCW\,58 can be expected to consist of amorphous silicates such as olivines (oxygen-rich magnesium iron silicates: (Mg,Fe)$_2$SiO$_4$) and pyroxenes (oxygen-deficient magnesium iron silicates: (Mg,Fe)SiO$_3$), together with metal oxides, in particular corundum (aluminium oxide: Al$_2$O$_3$) \citep{Tamanai2017}. In the absence of mid-infrared spectra for RCW\,58, we cannot confirm or rule out any of these dust species. However, oxygen-deficient amorphous silicates have a lower peak opacity and a lower near-IR absorptivity compared to oxygen-rich silicates \citep{Ossenkopf1992}, so if the dust were composed primarily of pyroxene a larger mass of dust would be required to model the IR
SED. Corundum has an emissivity peak in the 12$\mu$m spectral region \citep{Onaka1989} but, since Al is 10 times less abundant than Si, we anticipate the contribution of corundum to the far-IR SED to be much
lower than either olivine or pyroxene. Further analysis is beyond the scope of the present paper. Following Paper I, we select spherical, amorphous silicate grains (olivine) whose optical properties are included in {\sc cloudy}.

As part of our modelling procedure, we adjust the grain size range and dust-to-gas ratio until we fit the IR SED. \citet{Mathis1992} suggested two populations of grains were required to explain the IR observations of RCW\,58: one with sizes between 0.002 and 0.008~$\mu$m, and another with sizes between 0.005 and 0.05~$\mu$m located at a larger distance from the star. These findings were based on IRAS fluxes at 25, 60 and 100~$\mu$m, where the 100~$\mu$m flux was highly uncertain due to background variation. 

With {\sc cloudy} we can include single-size grains and also populations with resolved size bins between a minimum, $a_\mathrm{min}$, and maximum, $a_\mathrm{max}$, grain size with a power-law distribution. We assume that the grains have a MRN \citep{Mathis1977} power-law size distribution $N(a)\propto a^{-3.5}$ and use 10 bins for each resolved population. For each bin $i$, the ratio of maximum to minimum size $a_i/a_{i-1}$ is taken to be constant and equal to $(a_\mathrm{max}/a_\mathrm{min})^{1/10}$.

\subsection{Constraints on the models}
  \label{ssec:constr}

In order to assess if a model is a good fit to the observations, we
need to define some constraints. For this, we have selected the following:
\begin{enumerate}

\item the shape and flux of the IR photometry obtained from all images
  shown in Figure~\ref{fig:rcw58mw}, illustrated in Figure~3 and listed in
  Table~\ref{tab:fot};

\item the nebular free-free emission at $\nu_{0} = 2.735$~GHz obtained by ATCA. (Note that this is instead of the H$\alpha$ flux that we used in Paper I);

\item the gas electron temperature, $T_\mathrm{e}$, determined from the [N\,{\sc
    ii}]\,$\lambda\lambda$\,5755/6584\, emission-line ratio for an aperture corresponding to the Magellan Clay telescope observations of \citet{Esteban2016};

\item the optical (H$\alpha$) and IR radial surface brightness distributions. For this purpose we extracted surface-brightness profiles from the images shown in Figure~\ref{fig:rcw58mw} for regions corresponding to the LC and RC clumps using the {\sc cart2pol} routine\footnote{\url{https://github.com/e-champenois/cart2pol}}. The resulting profiles are shown in the top panels of Figure~\ref{fig:profiles}. 

\end{enumerate}

\section{Results}
\label{sec:results}

In this section we describe our {\sc cloudy} photoionization models that best reproduce the optical, IR and radio properties of regions LC and RC clumps in RCW\,58.
We performed preliminary calculations using a shell of pure gas with appropriate density and chemical abundances together with the stellar atmosphere model WNL\,06-14 described in \S~\ref{ssec:stell}. We found that this stellar atmosphere model has insufficient UV photons to produce the He\,{\sc i} and [N\,{\sc ii}] line strengths while at the same time exceeds that of the [S\,{\sc ii}] lines when compared to the spectroscopic observations of \citet{Esteban2016}.
For this reason, we also ran {\sc cloudy} models using atmosphere WNL\,06-13, which corresponds to a slightly larger transformed radius, $R_\mathrm{t}$ and has a higher helium-ionizing photon rate (see Table~\ref{tab:powr}).

Adding dust to the models increases the complexity. We found that multi-layer models were required to satisfy the constraints defined above (see \S~\ref{ssec:constr}). The locations and thicknesses of the different layers were guided by the H$\alpha$ and IR radial surface brightness profiles shown in Figure~\ref{fig:profiles}.

\begin{figure}
\centering
\includegraphics[width=0.5\textwidth]{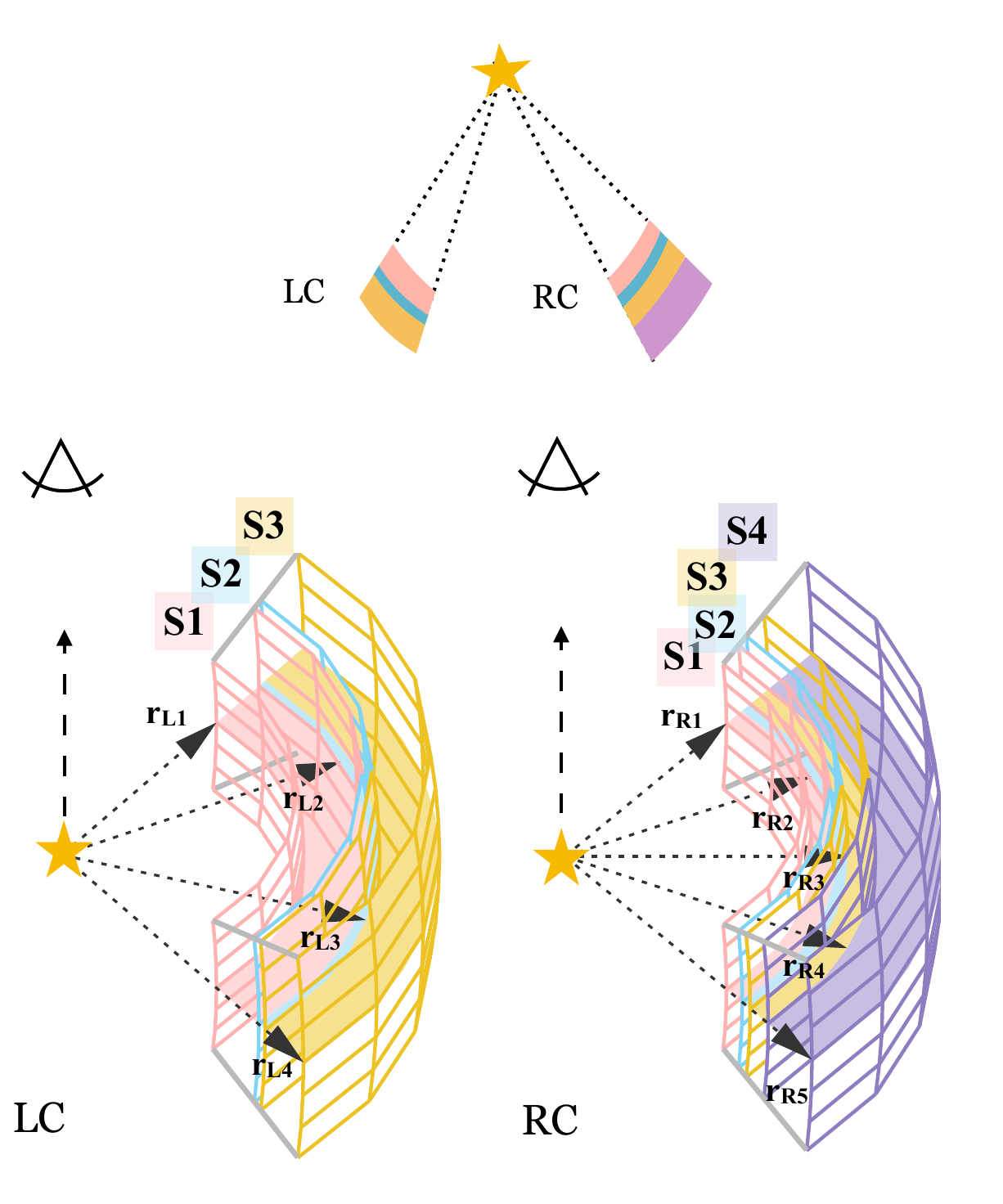}
\caption{Schematic view of the gas and dust distributions in the LC (left panel) and RC (right panel) clumps. The innermost (white) regions 
are gas and dust free. Regions S1 contain only gas, S2 contain small grains, while regions S3 and S4 contain a combination of distinct populations of small and large grains. The properties of each layer are described in Sections~4.1 and 4.2. The clumps are taken to be pieces of a thick ring viewed pole-on, as illustrated in the upper figure, and the line-of-sight depth is $\sim$5.5 times the plane-of-the-sky width. In the 3D views, the clumps have been tilted at 40$^\circ$ to the line-of-sight,  to aid visualization of the different regions.} 
\label{fig:scheme}
\end{figure}

From Figure~\ref{fig:profiles} we see that there are unambiguous differences in the dust distribution between LC and RC. In both regions, the peak of the {\it WISE} 12~$\mu$m profile coincides with that of the H$\alpha$ emission because this broad-band IR filter includes the contribution of emission lines from the ionized gas \citep[see, e.g.,][]{Mathis1992,Toala2015}. For LC, the peaks of the optical and the IR profiles are almost coincident. The 70 and 100\,$\mu$m IR profiles extracted from the {\it Herschel} PACS observations are only offset 6~arcsec behind that of the optical emission. The 22~$\mu$m \textit{WISE} emission peaks at the same position as the PACS profiles but is much broader due to the lower spatial resolution of these observations. On the other hand, the RC profiles suggest a more stratified distribution of dust. For example, the offset between the peak of the optical profile and those of the {\it Herschel} profiles is 21~arcsec, with the {\it WISE} 22~$\mu$m emission peaking at an intermediate position.

A schematic view of the gas and dust distributions for the LC and RC models is shown in Figure~\ref{fig:scheme}. 
The 3D geometry assumed for each  of LC and RC is that of a wedge-shaped sector of a 3D ring whose plane-of-the-sky cross section is an annulus sector with diagonal $w^2 = r_\mathrm{out}^2 - r_\mathrm{in}^2$, and whose line-of-sight depth is the chord length between the 2 points where the tangent to the inner radius $r_\mathrm{in}$ intersects the outer boundary at $r_\mathrm{out}$. Generally, the assumed line-of-sight depth is a few times ($\sim$5.5) the distance $r_\mathrm{out} - r_\mathrm{in}$.

Multiple regions were required to model LC and RC and their characteristics are described in detail below.
The fitting procedure is iterative and can be described as follows. From the surface brightness profiles we fix the inner and outer radii of the density distributions, which we take to be constant in each layer. For each layer of the LC and RC there are then eight free parameters: the hydrogen density $n$, filling factor $\epsilon$, and grains defined by size ranges $a_\mathrm{min}$, $a_\mathrm{max}$ for the two populations, the big-to-small grain population ratio $B/S$, and the dust-to-gas mass ratio $D/G$. The combination of $n$ and $\epsilon$ parameters are used to model the ionized gas emission, and together they determine the synthetic 12~$\mu$m and 2.7 GHz photometries, as well as the resultant emission lines. We choose values of $n$ from the range derived from the ATCA observations (see Section~\ref{ssec:ffe}) and the corresponding value of epsilon is that which is required to give a good fit to the 2.7~GHz and 12~$\mu$m photometric points.

Once the model hydrogen density and filling factor have been determined, we turn to the dust grains.
The big grain population is selected so that the model dust emission fits the peak of the IR part of the SED as well as the slope at longer wavelengths. This slope is determined solely by the large grain population $a_\mathrm{big}$. A population of small grains $a_\mathrm{small}$ is then needed to complete the shorter wavelength part of the SED and this is determined by adjusting its size range, and the $D/G$ and $B/G$ ratios.

In Appendix A we show the results of some experiments in the parameter space ($n_0$, $a_\mathrm{big}$: $a_\mathrm{min}$, $a_\mathrm{max}$) and how these key parameters are the ones which determine the model SED. 

\subsection{The LC model}

We found that 3 layers were required to account for the observational properties of LC. These regions were defined with reference to the surface brightness profiles of the top left panel of Figure~\ref{fig:profiles}. The gas densities are inspired by the ATCA results described in \S~\ref{ssec:ffe} and depicted in Figure~\ref{fig:den}. We report the results obtained using the stellar atmosphere model WNL~06-13. In the following, all radii are measured from the central star and converted to arcsec assuming a distance to the nebula of $d=3.83$~kpc.
\begin{enumerate}

\item S1 layer: $146~\mathrm{arcsec} < r < 186~\mathrm{arcsec}$.
This layer consists solely of gas with constant density $n=16$~cm$^{-3}$ and a filling factor $\epsilon = 0.03$.

\item S2 layer: $186~\mathrm{arcsec} < r < 192~\mathrm{arcsec}$. This intermediate layer comprises gas with constant density $n=31$~cm$^{-3}$ and a filling factor $\epsilon=0.08$ together with a population of small dust grains with a MRN power-law distribution with grain sizes between $a_\mathrm{min} = 0.002~\mu$m and $a_\mathrm{max} = 0.01~\mu$m. The dust-to-gas mass ratio (D/G) in this layer is $6.5\times 10^{-4}$.

\item S3 layer: $192~\mathrm{arcsec} < r < 228~\mathrm{arcsec}$. This outer layer is composed of gas with constant density $n=31$~cm$^{-3}$ and filling factor $\epsilon=0.08$, and two grain populations: the same small grain population as layer S2  and, in addition, a MRN power-law distribution of large grains with sizes between 0.6 and 0.9~$\mu$m. The relative proportion by mass of the large grain population to the small grain population (B/S) is 60. The total dust-to-gas mass ratio (D/G) in this layer is $4.0\times 10^{-2}$.

\end{enumerate}

\begin{table*}
  \begin{center}
  \caption[]{Input parameters and final masses for the different layers of the LC and RC models.}
     \begin{tabular}{lccccccc}
     \hline
     \hline
Parameter                              & \multicolumn{3}{c}{L 6-13} & \multicolumn{4}{c}{R 6-13}  \\  
                                       & S1     & S2         & S3          & S1   & S2        & S3          & S4\\
     \hline
Distance (kpc)                         & \multicolumn{3}{c}{3.8}      &\multicolumn{4}{c}{3.8} \\
$\log_{10}(L_{\star}/\mathrm{L_\odot})$& \multicolumn{3}{c}{5.91}            & \multicolumn{4}{c}{5.91}  \\
Inner radius[``]                        & 146   & 186         & 192         & 144 &158         & 165         & 181\\
Outer radius[``]                        & 186   & 192         & 228         & 158 &165         & 181         & 220\\
    \hline
$n_{\mathrm{0}}$\,[cm$^{-3}$]          & 16    & 31          & 31          & 63  & 63         & 63          & 63\\
Filling Factor ($\epsilon$)            & 0.030 & 0.080       & 0.080       &0.015& 0.015      &0.015        &0.015 \\
    \hline
$a_\mathrm{small}$ [$\mu$m]            & -     & 0.002-0.010 & 0.002-0.010 &  -   & 0.002-0.010 & 0.002-0.010 & 0.002-0.010\\
$a_\mathrm{big}$ [$\mu$m]              & -     & -           & 0.6-0.9     &  -   & -           & 0.6-0.9     & 0.6-0.9  \\
$B/S$                                  & -     & -           & 60          &  -   & -           & 80          & 127\\
$D/G$                                  & -     &8.269$\times 10^{-4}$ & 5.044$\times 10^{-2}$ &  -   &4.961$\times 10^{-4}$ & 4.019$\times 10^{-2}$   & 6.334$\times 10^{-2}$  \\
    \hline
Gas mass [M$_{\odot}$]               & 0.0027 & 0.0025      &0.0207      &0.02458 & 0.00831     & 0.03024     & 0.10266\\
Dust mass [$10^{-5}$M$_{\odot}$]     & -      & 0.2059      &104.6       &  -     & 0.4123      & 121.5       & 650.3\\
     \hline
     \end{tabular}
     \label{tab:mods}
  \end{center}
\end{table*}

Full details of this model, labelled L\,6-13, are listed in Table~\ref{tab:mods}. Taking into account all three layers, the total ionized gas mass of the 3D clump is 0.026~M$_{\odot}$,
with a dust mass of $M_\mathrm{D}=0.001$~M$_{\odot}$. Region S3 contains the largest proportion of dust, which reaches 5~per~cent of the mass of ionized gas in this layer. Assuming the 3D geometry described above, we calculated the predicted surface brightness profiles as a function of angular distance from the star of the different infrared bands obtained from this {\sc cloudy} model, and convolved them with the appropriate instrumental broadening. The results are shown in Figure~\ref{fig:profiles}, bottom left panel. The model predicts an angular separation between the H$\alpha$ peak and the {\it Herschel} wavelength bands of 6~arcsec, similar to the observations. In Figure~\ref{fig:profiles}, the observational surface brightness profiles (top left panel) for the longer wavelength emission are narrower than that of the H$\alpha$ emission. This suggests that the line-of-sight depth for the dusty material should be smaller than that of the pure gas component: we can imagine an dusty nucleus within a more extended clump of gas, for example.

Finally, in Figure~\ref{fig:temdmod} (left panel) we plot the dust temperature as a function of the radial distance from the star that results from the 1D spherically symmetric {\sc cloudy} model. The small grains have temperatures around 60~K while the large grains have temperatures around 30~K. The mass-weighted average dust temperature is represented by the heavy line in the figure and we can see that, in the layers with mixed grain populations, it is dominated by the temperature of the large grains, which comprise the bulk of the mass. The average dust temperature is around 30~K, which is considerably lower than the value of $\sim44$~K obtained from the MBB fit to the LC photometry (see \S~\ref{sec:obs}). This suggests that MBB parameters should be treated with caution when mixed grain populations are expected.

\subsection{The RC model}

The RC model requires 4 layers to adequately explain the observations. As for LC, the regions are defined with regard to the surface brightness profiles shown in Figure~\ref{fig:profiles} (top right panel) and the gas densities are taken from the ATCA results reported in \S~\ref{ssec:ffe}. Again, we used the {\sc powr} stellar atmosphere model WNL~06-13 in our {\sc cloudy} simulations.

\begin{enumerate}
    \item S1 layer: $144~\mathrm{arcsec} < r < 158~\mathrm{arcsec}$. This layer contains only gas with constant density $n=63$~cm$^{-3}$ and filling factor $\epsilon=0.015$.
    
    \item S2 layer: $158~\mathrm{arcsec} < r < 165~\mathrm{arcsec}$. This layer contains gas with constant density $n=63$~cm$^{-3}$ and filling factor $\epsilon=0.015$ together with a population of small dust grains with a MRN power-law distribution with grain sizes between $a_\mathrm{min} = 0.002~\mu$m and $a_\mathrm{max} = 0.01~\mu$m. This is the same dust population as layer S2 of the LC model. The dust-to-gas ratio (D/G) in this layer is $3.9\times10^{-4}$.
    
    \item S3 layer: $165~\mathrm{arcsec} < r< 181~\mathrm{arcsec}$. This region requires two grain populations: the same small grain population as layer S2 and a MRN power-law distribution of large grains with sizes between 0.6 and $0.9~\mu$m. The gas properties are the same as for layers S1 and S2 and the total dust-to-gas ratio is $3.15\times 10^{-2}$. The relative proportion of the big to the small grains (B/S) is 80.
    
    \item S4 layer: $181~\mathrm{arcsec} < r < 220~\mathrm{arcsec}$. A final, external layer is required with the same gas properties as the interior layers and the same two grain populations as layer S3. However, in layer S4 the ratio of big to small grains is much higher, equal to 127, and the total dust-to-gas mass ratio is $5.0\times 10^{-2}$ in this case.
\end{enumerate}

Table~\ref{tab:mods} lists the details of the R\,6-13 model including the $D/G$ and $B/S$ values. Taking into account all four layers, the total ionized gas mass of this model is 0.17~M$_{\odot}$ and the dust mass is $M_\mathrm{D}=0.016$~M$_{\odot}$. In region S3 the dust mass is 4~per~cent of the gas mass, while in region S4 this rises to 6~per~cent.  The synthetic surface brightness profiles for model R\,6-13 are shown  Figure~\ref{fig:profiles} bottom right panel. There are 3 main peaks in these profiles: one for the optical (gas) emission, one for the $22~\mu$m emission, and one for the longer wavelength emission. The model reproduces the peak separations seen in the observations (see Fig.~\ref{fig:profiles}) but not the width of the profiles. As for the L\,6-13 model, the wider H$\alpha$ and narrower dust-emission surface brightness profiles can be explained if the line-of-sight depth of the emitting components is different, with the dusty regions being contained within a larger gas clump.

\begin{figure*}
\centering
\includegraphics[width=0.49\textwidth]{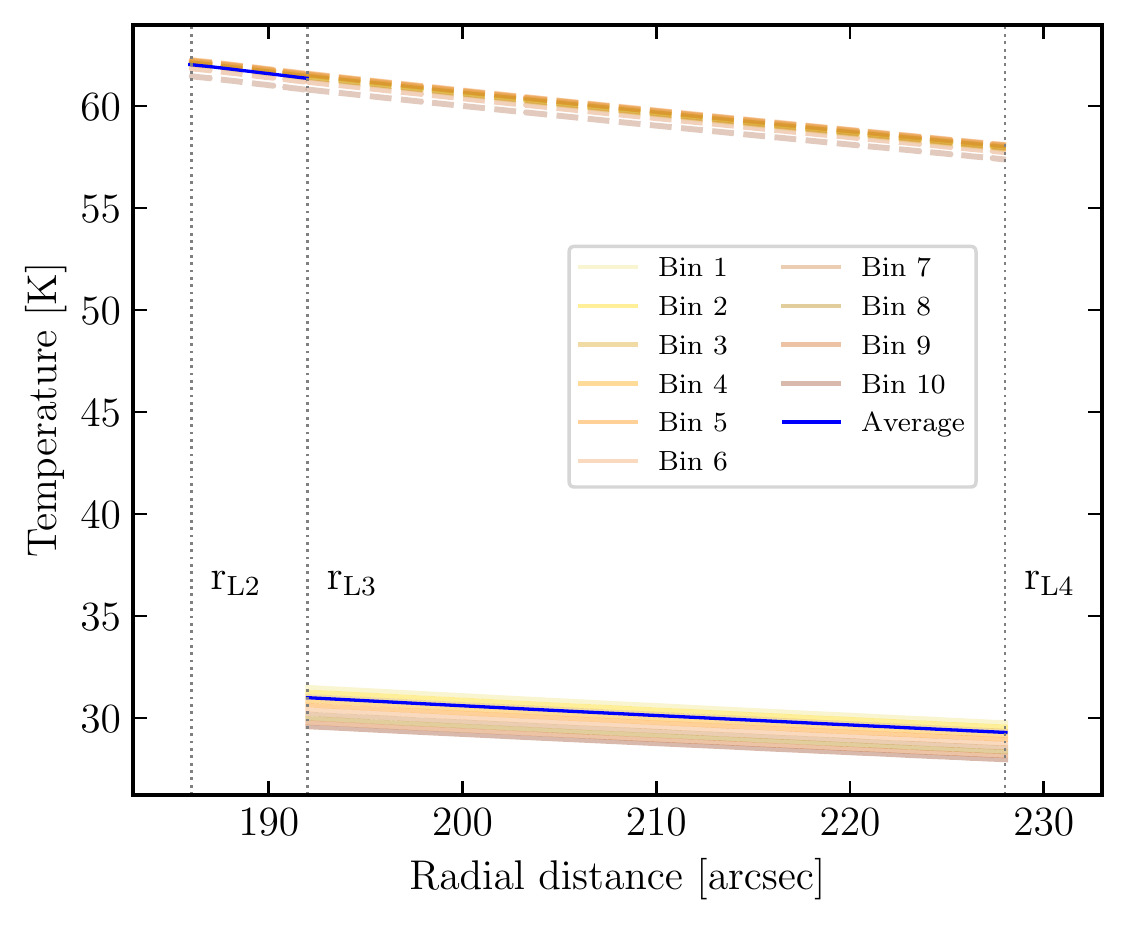}
\includegraphics[width=0.49\textwidth]{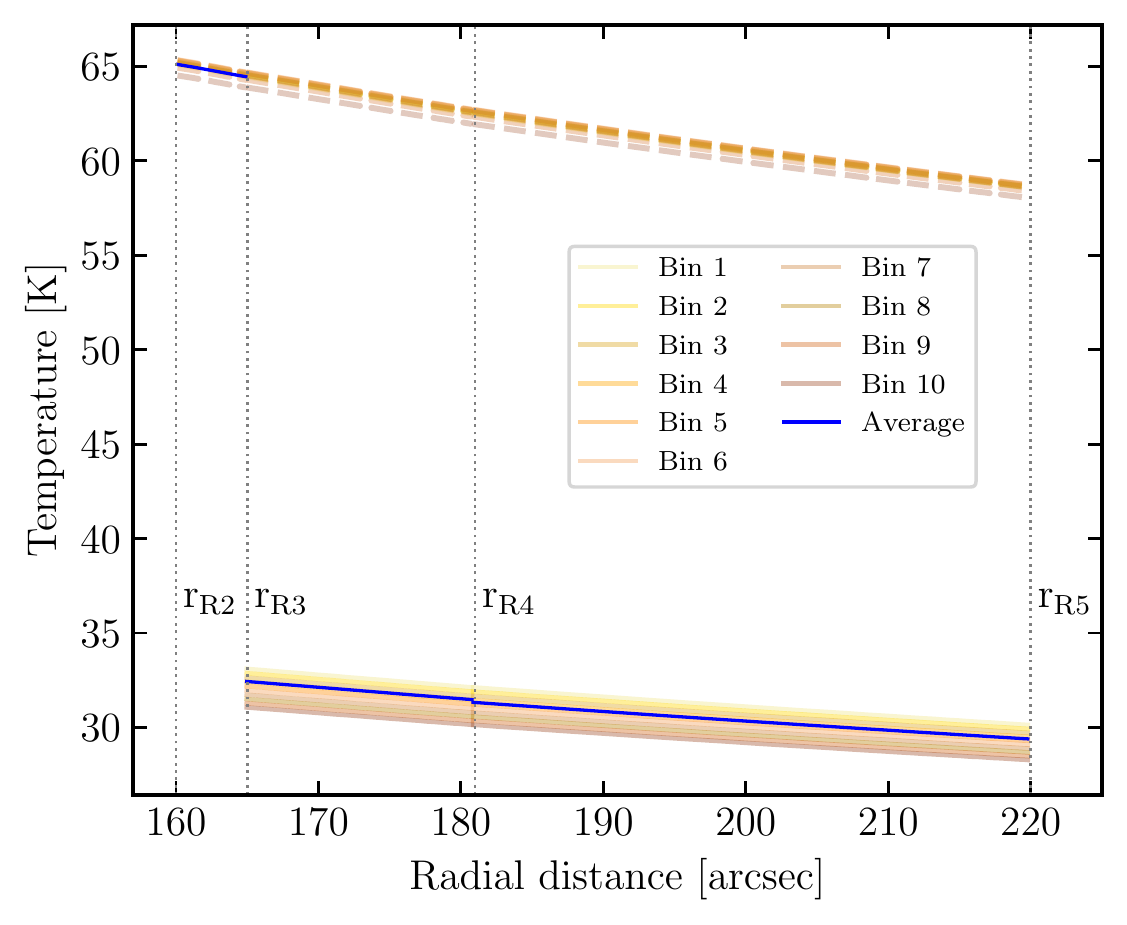}
\caption{Dust temperature as a function of radial distance from the central star for models L\,6-13 (left) and R\,6-13 (right). Different colours represent different bins in the dust grain size distribution. Solid (dashed) lines represent large (small) grains. Solid blue lines show the mass-weighted average temperature of the grains in each layer.}
\label{fig:temdmod}
\end{figure*}

In Figure~\ref{fig:temdmod} (right panel), we plot the dust temperature distribution as a function of radial distance from the star that results from the 1D spherically symmetric {\sc cloudy} model. Similarly to the L\,6-13 model, the R\,6-13 model discussed here resulted in small-grain temperatures around 60~K  and large-grain temperatures around 30~K. The mass-weighted average grain temperature is dominated by the large grains and is around 30~K. The grain temperatures are a little higher than in the L\,6-13 model because all the components are slightly closer to the star.

\subsection{Optical emission lines}
\label{ssec:elines}

We used the {\sc pycloudy} library of tools to make synthetic observations of optical emission lines from an aperture corresponding to the \citet{Esteban2016} observations. The observed spectrum is extracted from a small ($1\times8$~arcsec$^2$) region of the RC clump. In Table~\ref{tab:linesout} we list the emission-line fluxes from the spectroscopic observations of \citet{Esteban2016} together with the synthetic observations
obtained from our {\sc cloudy} models
using the two different {\sc powr} stellar atmosphere models, WNL\,06-14 and WNL\,06-13 \citep{Hamann2006}. RCW\,58 is a low-ionization nebula and the optical spectrum shows several He\,{\sc i} recombination lines.

Model R\,6-14 does a reasonable job at reproducing the general emission-line trends in the observed spectrum. The electron density, derived from the [O\,{\sc ii}]~3726,3729~\AA\, and [S\,{\sc ii}]~6716,6731~\AA\, doublets using {\sc pyneb} \citep{Luridiana2015} returns similar results to the value derived from the observations. However, the electron temperature, derived from the [N\,{\sc ii}] emission lines, is considerably lower. Moreover, the intensities of all the He\,{\sc i} recombination lines are lower than those observed, suggesting that He$^+$/He is lower in the model. This parameter depends on the ionizing source properties and is lower for softer ionizing radiation \citep{Delgado-Inglada2014}. Thus, we also ran {\sc cloudy} simulations using the stellar atmosphere model WNL\,06-13, which has a higher rate of Helium-ionizing photons and is a harder ionizing radiation source (see Table~\ref{tab:powr}).

\begin{table}
  \begin{center}
  \caption{Emission lines considered in our models. Observational data were taken from \citet{Esteban2016} and are relative to $\mathrm{H}\beta=100$.}
  \setlength{\tabcolsep}{0.5\tabcolsep}
     \begin{tabular}{lccc}
     \hline
     \hline
 Line                      & Observations      & R\,6-13    & R\,6-14 \\
\hline
    H$\beta$               & 100.0             & 100.0      & 100.0  \\
    H$\alpha$              & 310 $\pm$ 11      & 299.75     & 302.25 \\
    {[O\,{\sc ii}]} 3726   & 24.2 $\pm$ 1.2    & 23.83      & 12.71  \\
    {[O\,{\sc ii}]} 3728   & 33.5 $\pm$ 1.4    & 32.80      & 17.61  \\
    H\,{\sc i} 3734        & 2.19 $\pm$ 0.46   & 2.36       & 2.35  \\
    H\,{\sc i} 3750        & 4.14 $\pm$ 0.98   & 3.00       & 2.99  \\
    H\,{\sc i} 3770        & 3.88 $\pm$ 0.62   & 3.89       & 3.88   \\
    H\,{\sc i} 3797        & 3.59 $\pm$ 0.35   & 5.19       & 5.17   \\
    He\,{\sc i} 3819       & 1.75 $\pm$ 0.44   & 2.23       & 0.65  \\
    H\,{\sc i} 3835        & 6.70 $\pm$ 0.55   & 7.14       & 7.12  \\
    He\,{\sc i} 3888       & 31.3 $\pm$ 1.3    & 24.57      & 7.36 \\
    He\,{\sc i} 3964       & 3.2 $\pm$ 0.51    & 1.07       & 0.97 \\
    H\,{\sc i} 3970        & 15.77 $\pm$ 0.67  & 15.52      & 15.46 \\
    He\,{\sc i} 4026       & 4.80 $\pm$ 0.74   & 4.03       & 1.02 \\
    H\,{\sc i} 4101        & 30.75 $\pm$ 0.99  & 25.32      & 25.22\\
    H\,{\sc i} 4340        & 48.9 $\pm$ 1.2    & 46.07      & 45.94\\
    He\,{\sc i} 4387       & 1.0 $\pm$ 0.25    & 1.07       & 0.27 \\
    He\,{\sc i} 4471       & 8.84 $\pm$ 0.36   & 8.42       & 2.15 \\
    He\,{\sc i} 4921       & 2.21 $\pm$ 0.34   & 2.31       & 0.60 \\
    {[Fe\,{\sc iii}]} 4658 & 0.87 $\pm$ 0.22   & 1.02       & 0.65 \\
    {[O\,{\sc iii}]} 5007  & 14.14 $\pm$ 0.54  & 2.79       & 0.00 \\
    He\,{\sc i} 5015       & 4.46 $\pm$ 0.16   & 3.30       & 1.20 \\
    {[N\,{\sc ii}]} 5755   & 1.11 $\pm$ 0.12   & 0.68       & 0.49 \\
    {He\,{\sc i}} 5875     & 25.45 $\pm$ 0.90  & 22.99      & 5.87 \\
    {[N\,{\sc ii}]} 6548   & 101.3 $\pm$ 3.6   & 72.71      & 90.42\\
    {[N\,{\sc ii}]} 6584   & 312 $\pm$ 11      & 214.33     & 266.55\\
    He\,{\sc i} 6678       & 8.15 $\pm$ 0.37   & 6.47       & 1.68\\
    {[S\,{\sc ii}]} 6716   & 10.58 $\pm$ 0.47  & 11.02      & 25.40\\
    {[S\,{\sc ii}]} 6731   & 8.01 $\pm$ 0.36   & 8.25       & 18.88\\
    He\,{\sc i}  7065      & 3.99 $\pm$ 0.34   & 5.03       & 1.25\\
    {[Ar\,{\sc iii}]} 7135 & 4.66 $\pm$ 0.47   & 5.75       & 0.72\\
    {[S\,{\sc iii}]} 9068  & 19.2 $\pm$ 1.4    & 12.16      & 1.97\\
    {[S\,{\sc iii}]} 9530  & 59.9 $\pm$ 4.1    & 30.54      & 4.95\\
\hline
    $F$(H$\beta$) $\times10^{-14}$   &1.45$\pm$0.03  & 0.068    & 0.071  \\
    $T_\mathrm{e}$[N\,{\sc ii}]5755/6584   & 6370$\pm$220   & 6190     & 5450   \\
    $n_\mathrm{e}$[O\,{\sc ii}]3726/3729   & 80      & 87       & 80   \\
    $n_\mathrm{e}$[S\,{\sc ii}]6716/6731   & 90      & 80       & 72   \\
\hline
\end{tabular}
     \label{tab:linesout}
  \end{center}
\end{table}

The results of changing the stellar atmosphere model to WNL\,6-13 are presented in Table~\ref{tab:linesout}, labelled as R\,6-13. This model does an even better job of reproducing the general trends of the spectral lines and the He recombination line intensities are very similar to the observations. The low-ionization lines of  [O\,{\sc ii}] and [S\,{\sc ii}] are also a close match. The  electron temperature and density derived from the model emission lines are within the errors of those derived by the same methods using the observed line intensities.

It is interesting to note that none of our {\sc cloudy} photoionization models reproduces the [N\,{\sc ii}], [O\,{\sc iii}] 5007~\AA\, or [S\,{\sc iii}] emission-line intensities seen in the observations. We varied the abundance of N within the range of values suggested in Table~\ref{tab:abun} but this did not make a significant difference to the line intensities relative to H$\beta$. A slightly hotter star, as suggested by the temperature diagnostic, could explain the discrepancies between the observed [O\,{\sc iii}] and [S\,{\sc iii}] emission-line intensities and the model results. With regard to the [N\,{\sc ii}] results, we note that analysis of the full {\sc cloudy} model shows that [N\,{\sc ii}]/H$\alpha$ increases with radius from a minimum of 0.7 to a maximum at the outer radius of  1.7. For the small aperture on RC of the \citet{Esteban2016} observations, the [N\,{\sc ii}]/H$\alpha$ ratio is 1, while our R\,6-13 model finds 0.7 at this position. We remark that \citet{Smith1988} found that the [N\,{\sc ii}]/H$\alpha$ ratio varied between 0.9 and 2 along a slit with position angle $167^\circ$ crossing the shell roughly in the location of the region LC. We tentatively attribute the [N\,{\sc ii}] discrepancy to the line-of-sight depth taken through the model, or the positioning and spatial resolution of the model aperture. Returning to the higher ionization lines, another possibility is that there is a contribution from tenuous material behind the outer shock, which is propagating into a low-density surrounding medium (see the analysis presented by \citealp{Gruendl2000}). In optical images, the ejecta clumps are enveloped in a faint [O\,{\sc iii}] shell and some of this shocked gas emission will inevitably be included along the line-of-sight in the aperture.

We notethat the H$\beta$ surface brightness obtained by \citet{Esteban2016} is a factor $\sim$20 higher than that calculated by our models. We attribute this to small-scale density variations within RC, which are not accounted for in our model but which are emphasised in the small bright region covered by theaperture in theobservations. Including density profiles in our model would add complexity by increasing the number of free parameters. Because of this, we decided to give priority to the radio photometry obtained from ATCA observations, which is more representative of the average density properties of the clumps.

\subsection{Ionized gas mass estimation}

Assuming that RCW\,58 is optically thin and that its free-free emission is thermal, we can obtain a global estimation of the ionized mass of the nebula using the procedure outlined by \citet{Smith1970}. The mass of the nebula is given by
\begin{equation}
    M = (1 +3 a) M_\mathrm{H} n_\mathrm{e} V,
\end{equation}
\noindent where $M_\mathrm{H}$ is the mass of the hydrogen atom and $V$ is the effective radiating volume. \citet{Smith1970} considered that the nebula has uniform density and that the helium is all singly ionized, and found an expression for $V$ from the integrated emissivity at frequency $\nu_0$, 

\begin{equation}
    V = 50~T_\mathrm{e}^{0.35}~\nu_0^{0.1}~\frac{d^2 S}{n_\mathrm{e}}.
    \label{eq:smithv}
\end{equation}

\noindent where $S$ is the flux of RCW\,58 at $\nu_0$. The mass can then be expressed as

\begin{equation}
    \frac{M}{\mathrm{M}_{\odot}} = 1.24 \frac{(1+3a)}{n_\mathrm{e}} \left( \frac{S}{[\mathrm{Jy}]}\right) \left( \frac{d}{[\mathrm{Kpc}]}\right)^2  \left( \frac{\nu_0}{[\mathrm{Hz}]}\right)^{0.1} \left( \frac{T_\mathrm{e}}{\mathrm{K}}\right)^{0.35},
\end{equation}

\noindent where $a$ is the fraction of helium in the nebula. From optical data and our photoionization models we have $T_\mathrm{e}=6190$~K, $n_\mathrm{e}\mbox{[O\,{\sc ii}]}=87$~cm$^{-3}$ and an average value of the helium fraction is $a = 0.155$. Finally, we measured a flux of $4.383 \times 10^{-2}$~Jy of RCW\,58 from the ATCA observations and thus we estimate an ionized gas mass of 2.5~M$_{\odot}$.

\section{DISCUSSION}
\label{sec:disc}

We have constructed photoionization models using {\sc cloudy} for the two regions LC and RC in RCW\,58, which successfully reproduce the IR SED for both clumps. In addition, the same model reproduces the principal optical emission line intensities (relative to H$\beta$) and derived quantities such as the electron temperature and density for a small aperture on the RC clump.

In this section, we examine the dust physical properties and spatial distribution implied by our model and discuss our results in the context of the evolution of RCW\,58 and its central star, WR\,40. Furthermore, we compare our results for RCW\,58 with those we recently obtained for M\,1-67, another nebula around a WN8h star.

\subsection{Dust characteristics and spatial distribution}
\label{ssec:cidust}

Our best {\sc cloudy} models require the presence of two populations of grain sizes for both RC and LC. A population of small grains with sizes between 0.002 and 0.01~$\mu$m is needed to fit the short wavelength part of the SED, while a population of large grains with sizes between 0.6 and 0.9~$\mu$m is required to reproduce the peak and the long wavelength part of the SED. The small grains and gas absorb most of the UV radiation from the star but the large grains dominate the dust mass.

The two grain populations are not uniformly mixed in the gas. For both RC and LC the photoionization models require an inner layer devoid of dust (S1 in Fig.~\ref{fig:scheme}). Layer S2 contains only the population of small grains: for LC the dust mass in this layer is 0.08~per~cent of the gas mass, while for RC it is 0.05~per~cent (see Table~\ref{tab:mods}).  Layers S3 and S4 (if present) contain both small and large grains, with total dust mass rising to between 4 and 6~per~cent of the gas mass. The mass of the large grains is roughly two orders of magnitude greater than that of the small grains in these layers and is considerably higher in RC than in LC. We remark that the gas density in RC is double that of LC. 

\subsubsection{Grain sizes, shapes and chemical composition}
\label{sssec:sizes}

In our models, we have assumed that the grains are composed purely of silicates. This is because the envelopes of massive stars will be nitrogen rich and carbon poor as a result of the CNO cycle in the hydrogen burning core and the mixing that occurs in convective zones after the main sequence. Oxygen is therefore expected to dominate the dust chemistry. The total mass of silicate dust that can be produced is limited by the availability of the least abundant element that makes up the dust formula: in this case Si.
Our models require that~5 per cent of the total mass of the S3 and S4 layer material is in the form of silicate dust. Since each olivine molecule in the dust grains has a mean molecular weight of $\mu_\mathrm{mol}\approx 172$, then the dust molecule abundance relative to hydrogen can be found from the relation
\begin{equation}
\frac{n(\mathrm{dust})}{n(\mathrm{H})} = \frac{D}{G}\frac{1}{X \mu_\mathrm{mol}}
\end{equation}
where $D/G$ is the dust-to-gas mass ratio reported in Table~\ref{tab:mods} and $X$ is the hydrogen mass fraction of the gas. For the gas abundances used in this work (see Table~\ref{tab:abun}), the hydrogen mass fraction is $X\approx 0.6$. The grain molecule abundance is therefore $n(\mathrm{dust})/n(\mathrm{H}) \approx  5\times 10^{-4}$ and, since every grain molecule contains a silicon atom, we compare this with the Solar silicon abundance $n(\mathrm{Si})_\odot/n(\mathrm{H})_\odot \approx 4\times 10^{-5}$ \citep{gass2010}. Thus, the implication is that there are at 
least 10 times more silicon atoms bound up in grains in RCW\,58 than should be available, since the assumption is that no additional silicon has been formed by nucleosynthesis in WR\,40.

A possible explanation for this is because we use spherical grains in our models. Grains with larger far-IR absorption cross sections than spherical grains can require less mass to reproduce the observed IR SED. Spheroidal grains have far-IR absorption cross-sections larger by a factor of 1.5 to 3 than spherical grains with the same volume \citep{Siebenmorgen2014}. Also, the emission of spheroidal grains would be shifted to longer wavelengths \citep{Gomez2018}. 

Another possibility could be so-called {\it fluffy} grains (porous aggregates of grains), which have a greater absorption cross-section than compact grains with the same mass. For grains larger than $a > 0.01\mu$m, porosity develops by grain growth due to coagulation. To explore this possibility, we computed a {\sc cloudy} model using the same nebular parameters as
model L\,6-13 but changing to fluffy grains. We used grains comprised of 50~per~cent vacuum and the rest being silicate material, i.e. porosity of 0.5. This model, denoted LC(fluffy), had the same two populations of grain sizes as our standard model L\,6-13, that is, a population of small grains  with sizes between 0.002 and 0.010~$\mu$m, and a population of big grains with sizes between 0.5 and 0.9~$\mu$m. Model LC(fluffy) successfully reproduced the observed IR SED with the same spatial distribution of dust as that of L\,6-13. The model results in a total mass of dust of $M_\mathrm{LC(fluffy)} = 3.8 \times 10^{-4}$~M$_{\odot}$, of which 90~per~cent corresponds to the largest size grains. This is in contrast to the $10.5\times10^{-4}$~M$_\odot$ of dust required by the compact grain model. That is, fluffy grains can reduce the total amount of dust material required by a factor of 2 or more, depending on the porosity.

Although the results with fluffy grains seem promising, coagulation of small grains into porous large structures will only occur in cold, dense regions ($T_\mathrm{gas} < 100$~K and $n > 10^3$~cm$^{-3}$), where the grain relative velocities are small enough to allow the grains to stick to each other through low-velocity grain-grain collisions \citep{Dominik1997,Hirashita2009}. High velocity collisions will cause the grains to shatter.

Recent studies have demonstrated that the competing effects of coagulation and shattering create and maintain the porosity of the grains, and a maximum porosity of 0.7 to 0.9 is achieved for grain size around 0.1~$\mu$m, with larger grains 
being less porous \citep[see][]{Hirashita2021}. 
Our models imply that the dust mass is dominated by grain sizes much larger than 0.1$\mu$m and although the porosity of such grains is not optimum, it can still result in a factor 2 or 3 reduction in the total mass of grains required to explain the observations. 

We have not explored the effect of including other grain species, such as iron or amorphous carbon grains, on our results. This is because our underlying assumption is that the RCW\,58 nebula material was ejected from the envelope of the progenitor star, which we expect to be oxygen rich and carbon poor. However, radiative transfer modeling of the 2.4 to 670~$\mu$m spectral energy distribution of the dusty Homunculus Nebula around the Luminous Blue Variable star $\eta$ Carina suggests the presence of iron, pyroxene and other metal-rich silicates, corundum, and magnesium-iron sulphide \citep{Morris2017}. The total dust mass our models infer for RCW\,58 would be easier to explain if it were not so dependent on the abundance of Si.

\subsubsection{Multi-layer distribution of dust grains}
\label{sssec:distnm}

Our preferred {\sc cloudy} models require the grains to be stratified in different layers within the clumps. The part of the clump closest to the star is devoid of dust. There is a population of small grains, with sizes between 0.002 and 0.01~$\mu$m, which is found throughout the remainder of the clump, and a population of much larger grains, with sizes between 0.6 and 0.9~$\mu$m, which is found only in the part of the clump furthest from the star. Where the large grains exist, they dominate the dust mass, and in RC their proportion increases with distance from the star. We will discuss the following interpretations for the stratified dust distribution:

\begin{enumerate}
    \item The large grains and small grains did not all form in the same location and this different spatial origin is preserved in the current clumps. 
    \item The small grains result from the destruction of the large grains as a result of clump disruption by the stellar wind shock.
    \item The small grains and the large grains were well mixed until dynamical segregation occurred, due to the interaction of the clump with the stellar wind shock or due to radiation pressure on the large grains.
\end{enumerate}

The large grain sizes inferred by our models require extreme environments and special physical processes such as coagulation through grain-grain collisions. Large grain sizes are also derived for other objects such as nova ejecta and Type II SN. Recent high-sensitivity ALMA observations with angular resolution of 0.1~arcsec of the red supergiant VY CMa reveal the presence of massive dusty clumps on spatial scales 70-700~AU from the star \citep{Kaminski2019}. Radiative transfer modelling, including grains up to 1~$\mu$m, suggests an atypically low gas content for the clumps in order to reconcile the total envelope mass of the star. It was not possible to determine an origin for the dusty mass-loss episodes in VY~CMa or to determine whether there was a change in dust properties with distance from the star from these observations. 

On the other hand, numerical studies of dust formation in the ejecta of low-mass systems that underwent a common envelope event show that dust is initially formed in the outer regions of the gas distribution, but it is not the largest grain size achieved in the calculations \citep{Iaconi2020} .
 
The largest grain sizes ($\sim0.1-1$~$\mu$m) are formed at intermediate times in the simulations and are not located at the outer edge of the gas distribution. If a similar dust formation history occurred in RCW\,58, extra processes have created the observed spatial distribution of grain sizes.

The clumpy shell seen in H$\alpha$ and 2.75~GHz images of RCW\,58 (see Figs.~\ref{fig:rcw58mw} and \ref{fig:rcw58atca}) is most likely the result of the interaction of the stellar wind shock with a layer of clouds. There is a rich literature on the interaction between a single cloud with a shock or a wind \citep[see, e.g.,][]{Pittard2016,Goldsmith2018} but studies of multicloud systems are less common \citep{Aluzas2012MNRAS,BandaBarragan2020,BandaBarragan2021}. Important features of the multicloud scenario are the thickness of the cloud layer and the spatial separation  between the mass sources (i.e., the porosity). Not only do the clouds interact with the main shock but there are subsequent interactions within the shocked layer. Although the parameters (intercloud density and shock velocity) are not tuned for nebulae around massive stars, the general scenario portrayed in simulations such as those of \citet{BandaBarragan2021} can be appreciated.

Initially, the impact of the main shock on the cloud layer triggers reflected and refracted shocks. After a short-lived transient period where the forward (refracted) shock splits into multiple shocks with different speeds as it encounters clouds of different densities, it then transitions to a steady transmitted shock that traverses the cloud layer, heating and compressing it. Individual cloud cores interact with the transmitted shock, expand laterally and are accelerated downstream. Kelvin-Helmholtz instabilities at the cloud-intercloud interfaces strip material from the clouds and entrain it into the flow. When the forward shock exits the cloud layer, it re-accelerates, and low-density entrained material moving with the post-shock flow also expands in the stream-wise direction. Individual clouds are stretched and produce filamentary tails. \citet{BandaBarragan2021} show that, in radiative models, strong cooling can promote the clumping and fragmentation of dense gas. In RCW\,58 there is the added process of photoionization, which will prevent the flow from cooling below 6500~K. Turbulence and mixing occur in the low-density, heated flow behind the accelerating forward shock.

In the context of RCW\,58, we see that different parts of the nebula are at different stages of this process. Towards the east, the forward shock has clearly exited the cloud layer and has left behind elongated filaments. In the region of RC, and also towards the north east, the [O\,{\sc iii}] emission is not well-separated from the H$\alpha$ shell, suggesting that the forward shock is only just exiting the cloud layer \citep[see][]{Gruendl2000}. Large linewidths were found in the optical spectra by \citet{Smith1988} in the vicinity of LC, suggestive of turbulent gas motions. The radial velocity with respect to the local standard of rest found in these long-slit spectra is about 90~km~s$^{-1}$. In addition, blueshifted absorption components with $v_\mathrm{LSR}$ between $-150$~km~s$^{-1}$ and $-100$~km~s$^{-1}$ have been detected in UV and optical spectra of the central star, WR\,40 \citep{Smith1984}. In the radiative simulations of \citet{BandaBarragan2021}, the forward shock re-accelerates to at most 0.4 times the initial shock velocity when it exits the cloudy layer, i.e., $v_\mathrm{as} \sim 0.4 v_\mathrm{b}$, and the post-shock flow velocity is $v_\mathrm{psh} = 0.75 v_\mathrm{as}$ for a fast shock. If we take the range of observed velocities as the post-shock flow velocities, then the initial shock velocity would be $300 < v_\mathrm{b} < 500$~km~s$^{-1}$. 

What consequences does this scenario have for the dust in RCW\,58?  While traversing the cloud layer, the forward shock has a lower velocity than the original shock wave, $v_\mathrm{fs} = K v_\mathrm{b}/\chi^{1/2}$, where $\chi$ is the density contrast between the cloud and the intercloud gas, and $K\sim2$ \citep{BandaBarragan2020}. The transmitted shock velocity within the cloudy layer, assuming a density contrast of $\chi = 10^2 (10^3)$, will then be $v_\mathrm{fs} \sim 60(20)$~km~s$^{-1}$ for the lower value of $v_\mathrm{b}$ or $v_\mathrm{fs} \sim 100 (30)$~km~s$^{-1}$ for the higher value.

Grains can be destroyed by shock waves through thermal and non-thermal sputtering, and through grain shattering \citep[see, e.g.,][]{Jones1996}. Sputtering arises from gas-grain impacts and returns grain material to the gas phase. Thermal sputtering is only important for fast shocks ($v_\mathrm{s} > 150$~km~s$^{-1}$), and non-thermal sputtering is the main dust-destruction process for intermediate shock velocities ($50~\mbox{km~s}^{-1} < v_\mathrm{s} < 150$~km~s$^{-1}$). Grain shattering dominates below $v_\mathrm{s} < 50$~km~$^{-1}$. For cloud-intercloud density contrasts $\chi = 10^2$, the range of transmitted shock velocities we derived for the cloudy layer falls within the range for non-thermal sputtering to be important. However, the timescale for dust destruction by sputtering for gas densities $n \sim 100$~cm$^{-3}$ is of the order $10^6$~yrs, far larger than the age of the nebula.  Thus, sputtering is not expected to be important in RCW\,58. For higher density contrasts, $\chi = 10^3$, the transmitted shock velocities are outside the sputtering range. However, we can expect shattering to be an important grain destruction mechanism. Shattering conserves the total grain mass but alters the grain size distribution. Thus, shattering  could produce the population of small grains required by our {\sc cloudy} models. However, the layer S2, which contains only small grains, should have a similar dust-to-gas mass ratio as the other layers, S3 and S4, if shattering is responsible.

We now consider segregation of the grains into the populations of large grains and small grains with different spatial distribution that our models require. We assume that the two populations of grains are initially well-coupled to the gas in the clumps. When the shock encounters a clump, the gas is compressed but the grains, due to their large inertia, decouple from the gas. The acceleration of the grains is opposed mainly by  collisional  drag forces. Collisional drag is due to the direct collisions of grains with atoms and ions, it is proportional to the grain cross section and inversely proportional to the grain mass. Large and high-density grains are therefore least affected by drag and remain decoupled from the gas.

The large grains are least affected by the refracted and reflected shocks and the gas instabilities, so their inertia will cause them to follow the kinematics of the original multicloud layer. The small grains, which our models require to be very small (0.002 to 0.01~$\mu$m), will follow more closely the kinematics of the shocked gas. This behaviour has been shown qualitatively, albeit for much higher shock velocities, in recent simulations of dust survival in clumps passing through a supernova reverse shock \citep{Kirchschlager2019,Slavin2020}.  We postulate that in our models for LC and RC, the inner region S1 corresponds to the gas behind the principal reflected (i.e., reverse) shock, as does region S2, which contains gas and small grains with a normal dust-to-gas mass ratio. Regions S3 and S4, which are dominated by the large grain population and have a high dust-to-gas mass ratio, would then correspond to the original cloud layer and the high dust-to-gas ratio could be attributed to dense gas being stripped from clumps in this region while the large dust grains remain. It is possible that this stripped gas is now in the expanding, low-density, turbulent flow behind the accelerating forward shock.

Finally, we remark that gas-grain drift due to radiation pressure on dust grains could have played a r\^ole at early times, when the clumpy layer was much closer to the star. In photoionized regions, such as RCW\,58, Coulomb forces between gas ions (mainly protons) and charged dust grains lead to increased plasma drag forces, which couple the dust to the gas \citep{Draine2011}. However, the dust is subject to a much larger radiation force than the gas, which leads to a drift between the dust and gas components and even a total decoupling sufficiently close to the star. Large and small grains decouple and recouple at slightly different distances from the star \citep{Henney2019}, so this differential could have imprinted itself on the spatial distribution of the grain sizes soon after the clumps were ejected from the stellar envelope and the star became a source of ionizing photons. Note that in Paper~I, our model for M\,1-67 required two layers: one of pure gas and one in which populations of small and large grains were mixed.

\subsection{Tracing the origins of RCW\,58 and WR\,40}
\label{ssec:new}

Given the strong IR emission from the background of RCW\,58 we were not able to produce a complete model for the nebula. However, we can estimate the total mass (gas$+$dust) by extrapolating from our results tailored to LC and RC and, consequently, estimate the initial mass of the progenitor of WR\,40.

From Table~\ref{tab:mods}, the total masses estimated for the LC and RC are 0.027 and 0.173~M$_{\odot}$, respectively. Taking the average of these values as typical for the 23 clumps detected in the radio image of RCW\,58 we estimated $M_\mathrm{TOT}\approx2.3\pm1.7$~M$_{\odot}$, which is of the same order as the mass estimated from the ATCA observations (2.5~M$_\odot$, see Section~4.4) and consistent with previous determinations of $\lesssim$3~M$_\odot$ \citep{Smith1970,Chu1982}. These results confirm RCW\,58 as the least massive WR nebula \citep[see Paper~I and][and references therein]{Rubio2020}.

The current mass of WR\,40 has been estimated to be in the 26--28~M$_\odot$ range by \citet{Hamann2019}. Adding to this the mass of RCW\,58, and considering that massive stars lose up to $\sim$10~M$_\odot$ during the main sequence phase \citep[for stars with ZAMS masses above 30~M$_{\odot}$; see][]{Ekstrom2012,Georgy2012}, we can estimate the mass of the progenitor of RCW\,58. These results seem to suggest that WR\,40 evolved from a massive star with initial mass around $M_\mathrm{i} \approx 40^{+2}_{-3}$~M$_{\odot}$. 

Such small estimates of the total mass of RCW\,58 and the initial mass of WR\,40 argue against an LBV scenario, but the need for dust with grain sizes as large as 0.9~$\mu$m advocates for an eruptive formation scenario where $\dot{M} > 10^{-3}$~M$_{\odot}~\mathrm{yr}^{-1}$ \citep{Kochanek2011}.
An alternative formation mechanism for RCW\,58 is that of the common envelope (CE) scenario. The idea, initially proposed by \citet{Paczynski1967}, is that a companion in a close binary system accretes the H-rich envelope from a massive component. This causes the reduction of the period of the binary in a spiral-in process releasing orbital energy that then ejects the massive envelope of the primary, simultaneously creating a WR star and its associated nebula. A more extreme version of this scenario is {\it explosive CE ejection (ECEE)}, proposed by  \citet{Podsiadlowski2010}. In this mechanism, hydrogen-rich material from the secondary component is injected into the helium-burning shell of the primary component of a binary system. The energy released from explosive nucleosynthesis expels the H-rich envelope. This scenario could naturally explain the extra Si needed to fit the IR SED of RCW\,58 but would result in ejecta velocities far higher than those observed as well as requiring the primary to already be in a helium shell-burning stage of evolution (see Paper~I). We favour the orbital energy route for expulsion of the H-rich envelope, but are left with the conundrum of the large quantity of silicate grains implied by our results.

\subsection{Comparison with M\,1-67}
\label{ssec:cm167}

RCW\,58 and M\,1-67 share similar characteristics, besides harbouring WN8h stars. Their optical images are dominated by H$\alpha$ clumps and their thermal IR emission is produced by dust associated with these clumps. Neither nebula has undergone a strong interaction with the surrounding ISM and their optical spectra are characterised by low-ionization emission lines. However, RCW\,58 shows clear signs that the fast stellar wind from the central star has interacted with the circumstellar clumps, giving rise to the [O\,{\sc iii}] skin and leaving a turbulent flow in its wake. M\,1-67 appears not to have been overrun by a fast stellar wind, indeed, a morpho-kinematic model of M\,1-67 recently proposed by Zavala et al.\ (submitted) requires three concentric structures: an external toroidal structure, an intermediate structure composed of ballistically expanding clumps, and an inner hollow region being overrun by the fast wind of WR\,124.

In Paper~I we estimated the kinetic energy associated with the expanding clumps in M\,1-67 using the nebular mass calculated from our {\sc cloudy} models (9.2~M$_\odot$) and the measured expansion velocity of 46~km~s$^{-1}$ (see Paper~I and references therein), obtaining $E_\mathrm{K}\approx1.9\times10^{47}$~erg~s$^{-1}$. For RCW\,58, the expansion velocity is less well-defined, but adopting a value in the range 30--87~km~s$^{-1}$ \citep[see][]{Chu1982,Smith1988} together with a total mass of 2.5~M$_{\odot}$ (see Section~\ref{sec:results}),  we estimate the kinetic energy for RCW\,58 to be $E_\mathrm{K}\approx[0.3-1.9]\times10^{47}$~erg~s$^{-1}$. The similarity of these values suggests a similar origin for the ejected mass in each nebula. Moreover, the toroidal nature deduced for M\,1-67 appears to be replicated in RCW\,58, since our models require a ring of material seen approximately pole-on, rather than a complete shell \citep[see also][]{Chu1982}. This supports the suggestion that both M\,1-67 and RCW\,58 formed through a CE channel. If confirmed, these stars should be placed under scrutiny because they could evolve to become stripped-envelope stars, which could eventually be candidates for progenitors of gravitational wave sources \citep[see][]{Laplace2020}.

WR\,40 is the second WN8h star that we propose to have formed as the result of a CE channel. Another candidate is the runaway WN8h star WR\,16 and its associated nebula \citep{Marston1994}, which was recently studied by \citet{Cichowolski2020} through archival IR and new  Atacama Submillimeter Telescope Experiment (ASTE) molecular observations. Using simple fits to the IR photometry of the main shell around WR\,16, these authors estimated a dust temperature of 65~K. This star appears to have had a complex interaction with the surrounding ISM and \citet{Cichowolski2020} postulate that it has experienced an  LBV phase. We intend to perform more detailed modelling of this object in the future.

\section{SUMMARY}
\label{sec:summ}

We presented the analysis of the nebular and dust properties of RCW\,58 around the WN8h star WR\,40. The presence of extended IR emission along the line of sight hampers a global study of RCW\,58. For this reason, we selected two clumps located towards the south of the nebula as representative for the dust properties of RCW\,58, namely, LC and RC.

We used the photoionization code {\sc cloudy} to produce synthetic optical, IR and radio observations that were compared to publicly available data. Our models reproduced the derived nebular physical properties and the IR SED when using the  WNL\,06-13 stellar atmosphere model from {\sc powr}. Our models required RCW\,58 to be distributed into a ring-like structure (instead of a shell) with stratified gas and dust distributions. The total mass (gas$+$dust) in the RC and LC clumps are 0.027 and 0.173~M$_{\odot}$, respectively, with total $D/G$ ratios of $\sim$0.04 for both models.

Our {\sc cloudy} model suggested a stratification of dust grains, with the smallest grains closer to the star and larger grains preferentially at the far-side of the clumps. We thoroughly discuss the different possibilities for the origin of such a distribution. It is very likely that a combination of a physical phenomena, such as gas-grain decoupling and the greater inertia of large grains, as well as grain shattering, produces the required dust distribution. The high dust-to-gas mass ratio predicted by our models could be ameliorated if non-spherical grains are present or if grain compositions other than silicates are also taken into consideration.

We found that the total mass of RCW\,58 is $\approx2.3\pm1.7$~M$_{\odot}$, relatively small compared to other WR nebulae. As a result, we estimated that the initial mass of WR\,40 was $\approx40^{+2}_{-3}$~M$_\odot$. Such small values for the mass of RCW\,58 and predicted initial mass of WR\,40 argue against an LBV evolution. However, our {\sc cloudy} model requires dust as large as 0.9~$\mu$m, which can only form after an eruptive ejection of material with $\dot{M}>10^{-3}$~M$_\odot$~yr$^{-1}$.

We propose that RCW\,58 formed through a CE channel producing simultaneously the observed WR nebula and giving birth to WR\,40. This would make WR\,40 the second WN8h star, in addition to WR\,124,
to have been formed as a result of binary interaction, in particular a CE evolution. 
If this scenario is confirmed, WN8 systems are possible candidates for future progenitors of gravitational wave sources.

\section*{Acknowledgements}

The authors are indebted to C.E.\,Cappa for reducing the ATCA data of RCW\,58 and fruitful discussions in an earlier version of this project. They also thank R.\,Galv\'{a}n-Madrid (IRyA-UNAM) for helping them with the analysis of the ATCA observations. The authors would like to thank the referee for comments that improved the presentation of this paper. 
The authors are grateful for financial support provided by Direcci\'{o}n General de Asuntos del Personal Academico (DGAPA), Universidad Nacional Aut\'{o}noma de Mexico (UNAM), through grants Programa de Apoyo a Proyectos de Investigaci\'{o}n e Inovaci\'{o}n Tecnol\'{o}gica (PAPIIT) IA100720 and a IN107019. PJH also thanks Consejo Nacional de Ciencias y Tecnolog\'{i}a (CONACyT - Mexico) for a research student grant. JAT acknowledges support from the Marcos Moshinsky Foundation (Mexico). This work makes use of {\it Herschel}, and {\it WISE} IR observations. {\it Herschel} is an ESA space observatory with science instruments provided by European-led Principal Investigator consortia and with important participation from NASA. {\it WISE} is a joint project of the University of California (Los Angeles, USA) and the JPL/Caltech,
funded by NASA. This work has made extensive use of NASA’s Astrophysics Data System. 

\section*{Data availability}

The data underlying this article will be shared on reasonable 
request to the corresponding author.



\bibliographystyle{mnras}
\bibliography{ms} 

\appendix

\section{Exploration of parameter space}
\label{app:sp}
    
Here we present the exploration of the parameter space around the model described in Section~\ref{sec:results} for LC. The reduced $\chi^{2}$ of our best model was estimated to be 0.5599. We explored the effect on $\chi^{2}$ of varying the hydrogen density $n$, the filling factor $\epsilon$ and the size range 
of the big grains ($a_\mathrm{min}$ to $a_\mathrm{max}$). Model examples are listed in Table~\ref{tab:appendix}.

A first test was to reduce $n$ to half of the value used for the best model in accordance with the range of values obtained from the analysis of the ATCA observations (top row in Table~\ref{tab:appendix}). This model requires $\epsilon$ to increase by a factor of three to compensate for the mass. Retaining the same big grain size distribution, it was necessary to adjust $D/G$ and $B/S$ to fit the long wavelength part of the SED. The resultant fit was slightly worse than the best model.

Other test models, maintaining $n$ and $\epsilon$ fixed but now changing $a_\mathrm{min}$ and $a_\mathrm{max}$, were attempted and result in very similar fits to that of the model presented in the main text. $D/G$ ratios are unchanged but we note a change in the big-to-small ($B/S$) ratio of $\lesssim$10 per cent.

We note that small variations in the key parameters around our basic model produce very similar results but that our general conclusions are robust. 

\begin{table}
  \begin{center}
  \caption[]{Test models to show the validity of our best model for LC. The parameters are normalized to the values obtained from our best model described in Section~\ref{sec:results} and are labelled with the index 0.}
  \setlength{\tabcolsep}{0.5\tabcolsep}
     \begin{tabular}{ccccccccc}
     \hline
    Modified   & $n$       & $\epsilon$ & $a_\mathrm{min}$& $a_\mathrm{max}$ & S2 & S3 & $B/S$ &$\chi^2$\\
    parameter  &[$n_0$]    &[$\epsilon_0$]  & [$\mu$m]        & [$\mu$m]         &  [$D/G$]$_0$     & [$D/G$]$_0$  & &\\
     \hline
     $n$                                 &    0.5  &  3 & 0.6  & 0.9  & 0.6 &  0.7 & 1.11 & 0.6834\\
     $a_\mathrm{min}$                    &    1    &  1  & 0.5  & 0.9  & 1 &  1 & 0.93 & 0.5598\\
     $a_\mathrm{max}$                    &    1    &  1  & 0.6  & 1  & 1 &  1 & 1.07 & 0.5601\\
     $a_\mathrm{min}$, $a_\mathrm{max}$  &    1    &  1  & 0.5  & 1  & 1 &  1 & 1  & 0.5600\\
     
     \hline
     \end{tabular}
     \label{tab:appendix}
  \end{center}
\end{table}

\end{document}